%% file: RTTlowTRv2.tex
\documentclass{article}
\usepackage{amsfonts}
\usepackage{amsmath}
\usepackage{amssymb}
\usepackage{epsfig}
\usepackage{float}

\def\beq{\begin{equation}}
\def\eeq{\end{equation}}
\def\bq{\begin{quote}}
\def\eq{\end{quote}}

\def\simlt{\stackrel{<}{{}_\sim}}
\def\simgt{\stackrel{>}{{}_\sim}}

\input{macros.tex}
\begin{document}

\begin{titlepage}

\noindent
\begin{flushright}
%lr *** preprint number \\
\end{flushright}
\vspace{1cm}

\begin{center}
  \begin{Large}
    \begin{bf}
Neutralino and gravitino dark matter with low reheating temperature
        \end{bf}
  \end{Large}
\end{center}

\vspace{0.5cm}
\begin{center}
\begin{large}
Leszek Roszkowski,$^{a}$\footnote{On leave of absence from the
  University of Sheffield, U.K.}
 Sebastian Trojanowski,$^{a}$ Krzysztof Turzy\'nski$^{b}$
\end{large}

\vspace{0.3cm}
\begin{it}
${}^{a}$ National Centre for Nuclear Research, ul.~Ho\.za 69, 00-681, Warsaw, Poland \\
\vspace{0.1cm}
${}^{b}$Institute of Theoretical Physics, Faculty of Physics, University of Warsaw, \\ul.~Pasteura 5, 02-093, Warsaw, Poland\\
\end{it}

\vspace{1cm}
\end{center}

\begin{abstract}
  We examine a scenario in which the reheating temperature $T_R$ after
  inflation is so low that it is comparable to, or lower than, the
  freeze out temperature of ordinary WIMPs. In this case the relic
  abundance of dark matter is reduced, thus relaxing the impact of the
  usually strong constraint coming from the requirement that the
  universe does not overclose. We first re-examine the dynamics of
  freezeout during reheating. Next we apply a Bayesian approach to
  study the parameter space of the MSSM with ten free parameters, the
  Constrained MSSM and the singlino-dominated regions of the
  Next-to-MSSM. In each case we often find dramatic departures from
  the usually considered regime of high $T_R$, with important
  implications for direct detection dark matter searches. In
  particular, in the MSSM we examine WIMP mass range up to about
  5~TeV, and we find large regions of bino dark matter over the whole
  mass range, and of higgsino dark matter with mass over a similar
  range but starting from the $\sim1$~TeV value of the standard high
  $T_R$ scenario. We show that the prospects for bino detection
  strongly depend on $T_R$, while the higgsino is for the most part
  detectable by future one-tonne detectors. The wino, which is
  excluded in the standard scenario, becomes allowed again if its mass
  is roughly above $3.5$~TeV, and can also be partially detectable. In
  the CMSSM, the bino and higgsino mass ranges become much more
  constrained although detection prospects remain roughly similar. In
  the Next-to-MSSM we show that, at low enough $T_R$ wide ranges of
  singlino-dominated parameter space of the model become again
  cosmologically allowed, although detection prospects remain nearly
  hopeless. We also study the non-thermal contribution to the DM relic 
  density from direct and cascade decays of the inflaton. Finally, in 
  the framework of the MSSM we consider the case
  of a gravitino as dark matter. In this case we find strong bounds
  from overclosure and from Big Bang Nucleosynthesis, and derive lower
  limits on $T_R$ which depend on the gravitino mass and on the nature
  of the lightest ordinary superpartner.
\end{abstract}

\vspace{3cm}

\end{titlepage}

%%%%%%%%%%%%%%%%%%%%%%%%%

\tableofcontents

\section{Introduction}

In spite of persistent efforts of both experimenters and theorists,
the Standard Model (SM) still reigns supreme as a correct
phenomenological description of almost all data in particle physics.
However, the existence of dark matter (DM) offers one of a few
empirical hints pointing beyond the SM and suggesting that it has to
be incorporated into a more fundamental theory. A well-motivated
example of such a theory is the Minimal Supersymmetric Standard Model
(MSSM) (for a review see, e.g., \cite{Martin:1997}), which -- unlike
the SM -- offers a candidate for a DM particle. The most commonly
discussed case, the lightest neutralino, which is a mixture of the
fermionic superpartners of the gauge and Higgs bosons, represents a
weakly interacting massive particle (WIMP) and is stable if it is
the lightest supersymmetric particle (LSP).  Its relic abundance is
determined at so-called freeze-out, when the annihilations become
inefficient due to a decrease in its number density in the expanding
Universe and the production processes are already ineffective due to a
drop in the temperature of the primordial plasma (the Lee-Weinberg
scenario).   The abundance of two other well-motivated DM
candidates, a gravitino -- a fermionic partner of a graviton -- and an
axino -- a fermionic partner of an axion -- (see, e.g., recent review
\cite{Choi:2013}), if they are the LSP, is generated by scatterings
of the primordial plasma particles and from out-of-equilibrium decays
of the lightest ordinary supersymmetric particles (LOSP) which had
previously undergone freeze-out.

The questions of the origin and the properties of dark matter remain
among of the main driving forces of both experimental and theoretical
research in physics beyond SM. The latter activity includes both
performing increasingly accurate calculations of the DM detection
rates and relic abundance, including a critical reappraisal of the conditions in which this abundance
was determined.  The importance of this twofold approach becomes
obvious by noting that the evolution of the Universe has been
empirically tracked back to temperatures as high as
$\mathcal{O}(\mathrm{MeV})$, but to obtain an estimate for the DM
abundance one typically needs to make bold extrapolations to much
higher temperatures.

It is usually assumed that the early Universe underwent a period of
cosmological inflation during which an accelerated expansion of the
Universe was driven by the vacuum energy density of a scalar
field -- an inflaton.  After inflation the large potential energy of
the inflaton field was transformed into the kinetic energy of newly
produced particles in thermal and chemical equilibrium.  As a result
of this process, dubbed reheating, the Universe entered a radiation-dominated
(RD) phase, and its initial temperature $T_R$ is commonly called the
reheating temperature.\footnote{This synopsis contains unavoidable
  simplifications, as reheating is actually a gradual process. Nevertheless,
  one can still define the reheating temperature as the one corresponding
  to an effective conclusion of inflaton decays. }

Another commonly adopted assumption is that the scale of $T_R$ is much
higher than the mass scale  of the MSSM particles, which allows one to separate
the dynamics of reheating from that of  DM freeze-out.
Although this assumption is convenient, there is no
{\em a priori} reason that it has to hold in the early Universe.  Intriguingly, a
recent study \cite{Dai:2014jja} has found that in the most popular
models of large-field inflation $T_R$ may be required to lie within
one or two orders of magnitude from the electroweak scale if the value
of the spectral index is to remain very close to its observationally determined
central value.

In this paper we will explore the possibility that the reheating
temperature is comparable to the temperature of freeze-out, and will
investigate the ensuing implications for DM phenomenology relative to
the standard case. A number of analyses along these lines
have been performed before: in a generic case \cite{Giudice:2000}, as
well as in the context of the Constrained MSSM (CMSSM)
\cite{Fornengo:2002} and of more general supersymmetric models
\cite{Gelmini:2006Feb,Gelmini:2006May,Gelmini:2006Oct}. However, the
discovery of the Higgs boson with mass $m_h \simeq 126$~GeV
\cite{CMS:2012,ATLAS:2012}, together with negative results of the ATLAS
and CMS searches for supersymmetric particles with masses below $\sim
1$ TeV point towards the soft SUSY breaking mass scale \msusy\  at
least an order of magnitude larger than the \mz\ scale.  These results
imply a significant shift in the standard paradigm for supersymmetric
dark matter. Previously, from naturalness-based assumption of
$\msusy\lsim {\cal O}(1\tev)$ it followed that bino-like neutralino
was considered as the most natural and attractive candidate for the
WIMP \cite{Roszkowski:1991ng} in the MSSM, the choice which was also
most naturally realized in unified
models~\cite{Roberts:1993tx,Kane:1993td} for comparable ranges of
$\msusy$. However, with increasing values of \msusy\ the relic
abundance typically exceeds the observationally determined value of
$\Omega h^2=0.1199\pm0.0027$ \cite{Planck:2013} by orders of magnitude
already for DM mass of a few hundred GeV, unless special mechanisms
of resonant annihilations or so-called coannihilations are employed
\cite{Griest:1990kh,Gondolo:1990dk}.
On the other, in the case of a higgsino-dominated neutralino,
coannihilations are very effective \cite{Mizuta:1992qp}, 
and the relic abundance remains too low until its mass increases to $\simeq
1$~TeV, which, intriguingly, is the scale implied by LHC limits on \msusy\ and also by
the Higgs boson mass of $\simeq 126$~GeV. Interestingly, just such a
higgsino-like WIMP emerges in unified SUSY with a TeV scale of
\msusy\ \cite{Roszkowski:2009sm}. For wino-like neutralino the
cosmologically favored mass range is even higher, $\simeq
3$~TeV~\cite{Hisano:2006nn}.

Here we will show that the problem of DM overabundance can be
alleviated at low reheating temperatures. Hence this
assumption will lead to the opening up of previously cosmologically
disallowed regions in the WIMP parameter space. In particular, the
wino can again become experimentally allowed, a multi-TeV higgsino can
have a correct relic abundance, while the relic
abundance of the singlino can be reduced to an acceptable level, which
in the standard case is hard to achieve. Prospects for WIMP direct
detection can also be significantly affected.

The plan of the paper is as follows.  In Section~\ref{sec:fo}, we
briefly review the dynamics of freeze-out in order to set the stage
and to understand the impact of a low reheating temperature on a
cosmological evolution and on the relic WIMP abundance. In
Section~\ref{sec:neudm}, we employ the Bayesian approach to
investigate the parameter space of the MSSM, the CMSSM and the
Next-to-MSSM (NMSSM), and will identify the regions that are
phenomenologically acceptable, including producing the correct relic
density of the neutralino DM at low reheating temperature. 
In Section~\ref{sec:decays} we discuss and quantify the additional non-thermal contribution to the DM relic density from direct and cascade decays of the inflaton field to DM species and show that it can increase the DM relic density up to the measured value in otherwise underabundant scenarios.
In Section~\ref{sec:grav}, we extend the analysis to include the
gravitino and assume it to be the DM, taking into account bounds from
the Big Bang Nucleosynthesis (BBN) that inevitably arise in the
presence of a long-lived LOSP. 
We conclude in Section~\ref{sec:conc}.

\section{Dynamics of freeze-out}
\label{sec:fo}

In this section, we review the dynamics of freeze-out for high and low
reheating temperatures.

\subsection{High reheating temperature}
\label{sec:hightr}

An evaluation of freeze-out at high reheating temperatures has by now
become a standard textbook lore (see, e.g.,~\cite{Kolb:1990}).  One
assumes that the Universe was initially in the RD phase and that the
energy density of radiation with $g_{\ast}(T)$ effective degrees of freedom was given by $\rho_R =
(\pi^2/30)\,g_{\ast}(T)\,T^4$,  with the temperature $T$ inversely
proportional to the scale factor $a$, i.e.,~$T \sim a^{-1}$.  For some
stable particle species which are pair-annihilated into
radiation in equilibrium processes, the Boltzmann equations governing $\rho_R$ and the number
density $n$ of some relic species read:
%%%%%%%%%%%%%%
\begin{eqnarray}
\frac{d\rho_R}{dt} & = &  -4H\rho_R + 2\langle\sigma v\rangle\,\langle E\rangle\big(n^2-n_{\textrm{eq}}^2\big),\nonumber\\
\label{BoltzmannhighTR}
\frac{dn}{dt} & = &  -3Hn - \langle\sigma v\rangle\big(n^2 - n_{\textrm{eq}}^2\big),
\end{eqnarray}
%%%%%%%%%%%%%%
where $H$ is the Hubble parameter, $\langle\sigma v\rangle$ is
a thermally averaged annihilation cross-section times velocity for the
species, $\langle E \rangle $ is its average energy 
(which we approximate as $\sqrt{m_i^2 + 9T^2}$) and $n_{\textrm{eq}}$
is its equilibrium number density.

In the context of supersymmetric theories with the LSP being a DM
candidate this description should, in principle, be generalized by
considering a separate Boltzmann equation for each supersymmetric particle species
that is heavier than the LSP. Owing to $R$ parity, these states pair-
and co-annihilate and their decay chains all end up with the LSP.
Fortunately, it was shown in \cite{Gondolo:1990dk} that in this case the evolution of the Universe can still be
effectively described by a system of
equations~(\ref{BoltzmannhighTR}), if one replaces the number density
of a single particle species by $n=\sum_{i}n_i$, where the index $i$
runs over all the particle species, each with a number density $n_i$, and
$\langle \sigma v \rangle$ is replaced by
%%%%%%%%%%%%%%
\begin{equation}
\langle\sigma v\rangle_{\textrm{eff}} = \sum_{i=1}^{N}\sum_{j=1}^{N}{\langle\sigma_{ij}v_{ij}\rangle\,\frac{n_{\textrm{eq},i}}{n_{\textrm{eq}}}\frac{n_{\textrm{eq},j}}{n_{\textrm{eq}}}},
\label{eq:sigeff}
\end{equation}
%%%%%%%%%%%%%%
where $n_{\textrm{eq},i}$ stands for the equilibrium number density of $i$-th
particle species, $n_{\textrm{eq}} = \sum_i{n_{\textrm{eq},i}}$
and $\langle\sigma_{ij}v_{ij}\rangle$ stands for a thermally averaged
(co)annihilation rate for $i$th and $j$th  particle species (for
a detailed discussion see, e.g.,~\cite{Gondolo:1990dk,Edsjo:1997}).
The effective average energy released in the (co)annihilations of
relic species is given by
%%%%%%%%%%%%%%
\begin{equation}
  \langle\sigma v\rangle_{\textrm{eff}}\,\langle
  E\rangle_{\textrm{eff}} = \sum_{i=1}\sum_{j=1}{\big(\langle
    E_i\rangle + \langle
    E_j\rangle\big)\,\langle\sigma_{ij}v_{ij}\rangle\,\frac{n_{\textrm{eq},i}}{n_{\textrm{eq}}}\frac{n_{\textrm{eq},j}}{n_{\textrm{eq}}}}\,
  . 
\end{equation}
%%%%%%%%%%%%%%
This approach is sufficient for accurate determination of the DM
abundance, since, due to aforementioned chain decays, already before
freeze-out $n$ becomes the number density of the single stable
species, the LSP.

Having justified using the formalism of a single  particle
species in the case of frameworks with many states, like the
MSSM, we can now briefly describe the dynamics of freeze-out.
Eqs.~(\ref{BoltzmannhighTR}) can be approximately solved under
assumption that 
%%%%%%%%%%%%%%
\begin{equation}
\langle \sigma v\rangle =\left(\alpha_s +
  (T/m_\chi)\,\alpha_p\right)/m_\chi^2\,,
  \label{eq:ov}
 \end{equation}
%%%%%%%%%%%%%%
where $m_\chi$ is the WIMP mass. Before freeze-out WIMPs undergo
(co)annihilations but are also produced in inverse processes and remain in thermal
equilibrium.  These processes are efficient until (co)annihilation
rate remains larger than the expansion rate of the Universe,
i.e.,~$n_{\rm{eq}}\,\langle\sigma v\rangle > H\sim T^2/M_{\rm{Pl}}$, where
$M_{\rm{Pl}}$ is the Planck mass.  The freeze-out temperature $T_{\rm{fo}}$
below which this relation is no longer satisfied marks the onset of an
era where the DM number density changes only due to the expansion
of the Universe.  The present DM density calculated
from~(\ref{BoltzmannhighTR}) is therefore given by
%%%%%%%%%%%%%%
\begin{equation}
%\Omega_{\textrm{DM}}h^2(\textrm{high }T_R) \simeq \frac{4\sqrt{5}}{\sqrt{\pi}}\,\frac{\Omega_Rh^2}{T_0\,M_{\rm{Pl}}}\frac{1}{g_{\ast}(T_{\rm{fo}})} \,\frac{1}{m_{\chi}^{-2}\,(\alpha_s\,x_{\rm{fo}}^{-1} + \alpha_p\,x_{\rm{fo}}^{-2}/2)}\,\rm{GeV}^{-2}\,,
\Omega_{\textrm{DM}}h^2(\textrm{high }T_R) \simeq
\frac{2\sqrt{5}}{\pi\sqrt{2}}\,\frac{\Omega_Rh^2}{T_0\,M_{\rm{Pl}}}\frac{1}{g_{\ast}(T_{\rm{fo}})}
\,\frac{1}{m_{\chi}^{-2}\,(\alpha_s\,x_{\rm{fo}}^{-1} +
  \alpha_p\,x_{\rm{fo}}^{-2}/2)}\,\rm{GeV}^{-2}\,, 
\end{equation}
%%%%%%%%%%%%%%
where $T_0$ is the present temperature of the Universe, $\Omega_Rh^2$
is the radiation relic density and $x_{\rm{fo}} = m_\chi/T_{\rm{fo}}$
satisfies
%%%%%%%%%%%%%%
\begin{equation}
x_{\rm{fo}} = \ln\Big[\frac{3\sqrt{5}}{2\pi^{5/2}}\,\frac{g}{g_{\ast}(T_{\rm{fo}})}\,\frac{M_{\rm{Pl}}}{m_{\chi}}\,\big(\alpha_s\,x_{\rm{fo}}^{1/2} + 2\alpha_p\,x_{\rm{fo}}^{-1/2}\big)\Big]\,,
\label{xFhighTR}
\end{equation}
%%%%%%%%%%%%%%
where $g$ is the number of degrees of freedom of the DM.

The parameter $x_{\rm{fo}}$ depends very weakly on the details of DM
interactions.  Therefore,
%%%%%%%%%%%%%%
\begin{equation}
\Omega_{\textrm{DM}}h^2(\textrm{high }T_R) \sim \frac{1}{\langle\sigma
  v\rangle_{\textrm{fo}}} \, ,
\label{Oh2highTR}
\end{equation}
%%%%%%%%%%%%%%
where the subscript ``$\mathrm{fo}$'' corresponds to the value at
$T_\mathrm{fo}$ and we used (\ref{eq:ov}).  This approximation remains valid also for DM
relic density when $\langle\sigma v\rangle_{\textrm{fo}}$ is replaced
by $\langle\sigma v\rangle_{\textrm{eff,fo}}$.

\subsection{Low reheating temperature}

If the reheating temperature is comparable to the freeze-out
temperature, WIMPs may freeze out before the inflaton field has fully
decayed, i.e.,~when the energy density of the Universe is still
dominated by the energy density $\rho_\phi$ of the inflaton.
Therefore, the system of Boltzmann equations~(\ref{BoltzmannhighTR})
has to be extended to accommodate the decaying inflaton field
\cite{Giudice:2000}.  At the beginning of the reheating period the
temperature of the Universe rapidly increases from $T\approx 0$ to
some maximum value $T_{\textrm{max}}$ due to the inflaton decaying to
radiation.\footnote{The value of $T_{\textrm{max}}$ does not play a
  role in the determination of the DM relic abundance, since
  $\Omega_{\textrm{DM}} h^2$ is set mainly by the the rate of
  (co)annihilation processes near freeze-out. Other possible sources
  of DM are direct and cascade decays of the inflaton field and
  inelastic scatterings of the inflaton decay products
  \cite{Allahverdi:2002pu,Harigaya:2014waa}. We shall mention then
  only briefly at the end of our study, as they are model dependent
  and, moreover, in scenarios considered here the freeze-out
  temperature is very close to the reheating temperature, which, in
  principle, allows thermalization of DM. A recent discussion of
    these issues can be also found in \cite{Drewes:2014pfa}.}  At
this temperature -- though radiation is still being effectively
produced in inflaton decays -- the effect of the additional dilution
caused by the increased expansion of the Universe begins to dominate
and the temperature starts to decrease with an increasing scale
factor, scaling as $T \sim a^{-3/8}$. In other words, the same drop in
the temperature corresponds to a faster expansion of the Universe
during the reheating period than in the RD epoch.

The set of Boltzmann equations now reads:
%%%%%%%%%%%%%%
\begin{eqnarray}
\frac{d\rho_{\phi}}{dt} & = & -3H\rho_{\phi} - \Gamma_{\phi}\rho_{\phi},\nonumber\\
\label{Boltzmann}
\frac{d\rho_R}{dt} & = &  -4H\rho_R + \Gamma_{\phi}\rho_{\phi} + 2\langle\sigma v\rangle_{\textrm{eff}} \,\langle E\rangle_{\textrm{eff}}\big(n^2-n_{\textrm{eq}}^2\big),\\
\frac{dn}{dt} & = &  -3Hn - \langle\sigma v\rangle_{\textrm{eff}}\big(n^2 - n_{\textrm{eq}}^2\big),\nonumber
\end{eqnarray}
%%%%%%%%%%%%%%
where $\Gamma_{\phi}$ is the inflaton decay rate.

%%%%%%%%%%%%%%
\begin{figure}[t]
\begin{center}
\includegraphics*[width=9.5cm,height=8.5cm]{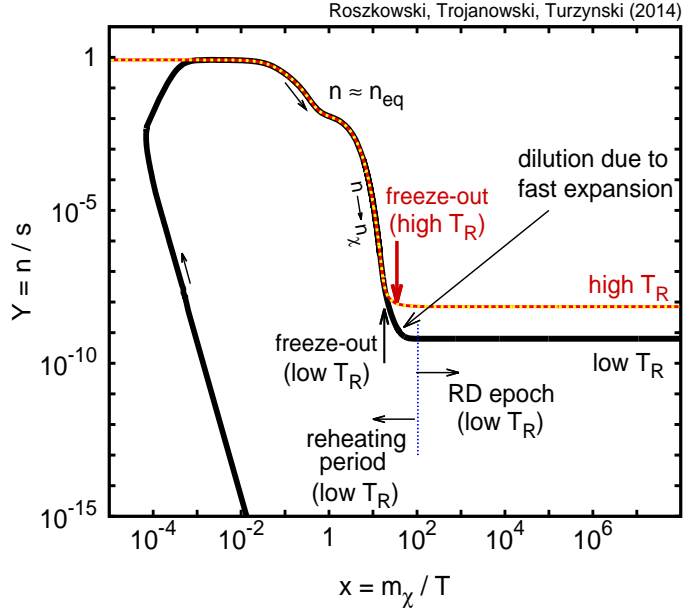}
\hspace{0.5cm}
\end{center}
\caption{Total yield 
$Y = n/s$ as a function of $x =  m_{\chi}/T$ 
in scenarios with low and high reheating temperature. A solid (dotted)
curve corresponds to the low (high) $T_R$ scenario. The beginning of
the RD epoch for the low $T_R$ scenario is denoted by vertical dotted
blue line.}
\label{yield}
\end{figure}
%%%%%%%%%%%%%%

The faster expansion during reheating is driven by the entropy production due to inflaton
decays and it continues until the inflaton decays completely.  One
conventionally associates the end of the reheating period with the
reheating temperature $T_R$ defined as the temperature of the Universe assuming that the inflaton decayed instantaneously,\footnote{In the reheating scenarios considered here,
  at $T_R$ given by eq.~(\ref{TRdef}) the Universe is typically still
  dominated by the inflaton field \cite{Giudice:2000} and the
  radiation-dominated epoch actually starts at a somewhat lower
  temperature.}  at the time corresponding to $\Gamma_{\phi}=H$,
%%%%%%%%%%%%%%
\begin{equation}
\Gamma_{\phi} = \sqrt{\frac{\pi^2\,g_{\ast}(T_R)}{90}}\,\frac{T_R^2}{M_{\textrm{Pl}}}.
%\hspace{2cm}\textrm{definition of }T_R.
\label{TRdef}
\end{equation}
%%%%%%%%%%%%%%
The reheating temperature $T_R$ is {\it a priori} unrelated to the
freeze-out temperature $T_{\rm{fo}}$ defined in Section
\ref{sec:hightr}.
In Figure~\ref{yield} we illustrate, in the
context of the MSSM, the temperature dependence of yield $Y$ defined as
%%%%%%%%%%%%%%
\begin{equation}
Y = \frac{n}{s},\hspace{1cm}\textrm{with    }s\, = g_\ast(T)\,\frac{2\pi^2}{45}\,T^3,
\end{equation}
%%%%%%%%%%%%%%
both for $T_R\gg T_{\rm{fo}}$ (high $T_R$ scenario) and $T_R\simlt
T_{\rm{fo}}$ (low $T_R$ scenario).  The solid (dotted) curve
represents the low $T_R$ (high $T_R$) scenario and supersymmetric mass
spectra have been selected in such a way that both number densities
reach their equilibrium values. Due to a faster expansion of the
Universe, for low $T_R$ the freeze-out occurs slightly earlier, with
typical $x_{\rm{fo}} = 10 - 25$, than for high $T_R$ where it
typically lies between 20 and 25.  If the decay of
the inflaton  stopped at $T_{\rm{fo}}$, the DM abundance would
be higher in the low $T_R$ scenario. However,  a continuous entropy
production keeps diluting it  until the reheating temperature is reached.
The end result is an overall reduction,\footnote{In principle one might expect
  a slight increase of the DM relic density, if freeze-out occurred just
  at the end of reheating period, since then the dilution period would
  not be present. However, we found that the maximum increase is at
  best a few percent, i.e., of the order of the error associated with
  this type of calculations.} of the DM abundance relative to high
$T_R$ scenarios \cite{Giudice:2000}.
 
Assuming again (\ref{eq:ov}), an approximate DM abundance resulting
from the set of Boltzmann equations (\ref{Boltzmann}) reads
\cite{Giudice:2000}  
%%%%%%%%%%%%%%
\begin{equation}
%\Omega_{\rm{DM}}h^2(\textrm{low }T_R) = \frac{5\sqrt{5}}{4\sqrt{\pi}}\,\frac{\Omega_Rh^2}{T_0\,M_{\rm{Pl}}}\,\frac{g_{\ast}^{1/2}(T_R)}{g_{\ast}(T_{\rm{fo}})} \frac{T_R^3}{m_{\chi}\,(\alpha_s\,x_{\rm{fo}}^{-4} + \frac{4}{5}\alpha_px_{\rm{fo}}^{-5})}\,\rm{GeV}^{-2},
\Omega_{\rm{DM}}h^2 = \frac{5\sqrt{5}}{8\pi\sqrt{2}}\,\frac{\Omega_Rh^2}{T_0\,M_{\rm{Pl}}}\,\frac{g_{\ast}^{1/2}(T_R)}{g_{\ast}(T_{\rm{fo}})} \frac{T_R^3}{m_{\chi}\,(\alpha_s\,x_{\rm{fo}}^{-4} + \frac{4}{5}\alpha_px_{\rm{fo}}^{-5})}\,\rm{GeV}^{-2},
%\label{Oh2lowTR}
\end{equation}
%%%%%%%%%%%%%%
%%%%%%%%%%%%%%
\begin{equation}
x_{\rm{fo}} =\ln\Big[\frac{3}{\sqrt{5}\pi^{5/2}}\,\frac{g\,g_{\ast}^{1/2}(T_R)}{g_{\ast}(T_{\rm{fo}})}\,\frac{M_{Pl}\,T_R^2}{m_{\chi}^3}(\alpha_s\,x_{\rm{fo}}^{5/2}+\frac{5}{4}\alpha_p\,x_{\rm{fo}}^{3/2})\Big].
\label{xFlowTR}
\end{equation}
%%%%%%%%%%%%%%
Finally we obtain
%%%%%%%%%%%%%%
\begin{equation}
\Omega_{\textrm{DM}}h^2 \sim \frac{1}{\langle\sigma v\rangle_{\textrm{fo}}}\,\frac{T_R^3}{m_{\chi}^3}\, ,
\label{Oh2lowTR}
\end{equation}
%%%%%%%%%%%%%%
where, similarly to (\ref{Oh2highTR}), the subscript
``$\mathrm{fo}$'' corresponds to the value at $T_\mathrm{fo}$ given by
(\ref{xFlowTR}), which is slightly larger than the value of the
freeze-out temperature obtained in the high $T_R$ scenario.  Of
course, in a full MSSM calculation one has to replace $\langle\sigma
v\rangle_{\textrm{fo}}$ with $\langle\sigma
v\rangle_{\textrm{eff,fo}}$ given by (\ref{eq:sigeff}).

\subsection{A comparison of the scenarios with a high and a low reheating temperatures}

As shown in eqs.~(\ref{Oh2highTR}) and (\ref{Oh2lowTR}), the DM relic
abundance in scenarios with high and low $T_R$ is determined by the
value of $\langle\sigma v\rangle_{\textrm{eff}}$ at the respective
freeze-out temperatures.  Since the freeze-out temperatures are very
similar in both cases, the following approximate relation holds:
%%%%%%%%%%%%%%
\begin{equation}
\Omega_{\textrm{DM}}h^2(\textrm{high }T_R) \simeq 
\left(\frac{m_{\chi}}{T_R}\right)^{3} 
%lr \left(\frac{m_{\chi}}{T_R}\right)^3 
\,\left( \frac{T_{\mathrm{fo}}}{m_{\chi}}\right)^3 \,\Omega_{\textrm{DM}}h^2 \, ,
\label{approxOh2mDM}
\end{equation}
%%%%%%%%%%%%%%
with $(T_\mathrm{fo}/m_\chi)^3$ factored out since its value changes
only in a narrow range.  From (\ref{approxOh2mDM}) it immediately
follows that in scenarios with low reheating temperatures,
$T_R<T_\mathrm{fo}$, the DM relic abundance is suppressed with respect
to scenarios with high reheating temperatures. Since the latter case
has been extensively studied and the DM relic density can be
easily calculated for a given WIMP type and mass, it is useful to
rephrase (\ref{approxOh2mDM}) in the following way. If
$\Omega_{\textrm{DM}}h^2$ is fixed at the observed
value of 0.12, a phenomenologically acceptable scenario is the one
where the standard prediction for
$\Omega_{\textrm{DM}}h^2(\textrm{high }T_R) $ is larger than the
observed value by a factor of
$(m_\chi/T_R)^3(T_\mathrm{fo}/m_\chi)^3$. In other words, SUSY configurations
which would be otherwise rejected as giving too large relic density
become acceptable at low reheating
temperatures. We shall explore this effect in Section
\ref{sec:neudm} when scanning a parameter space of some specific
SUSY models below.

%%%%%%%%%%%%%%
\begin{figure}[t]
\begin{center}
\includegraphics*[width=9.5cm,height=8.5cm]{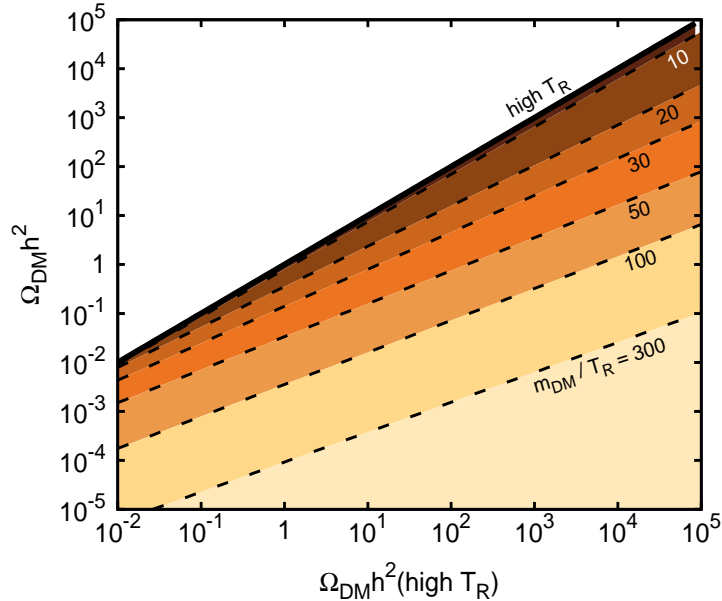}
\end{center}
\caption{A relationship between the relic density of DM
  $\Omega_{\textrm{DM}}h^2$ in low $T_R$ scenarios and
  $\Omega_{\rm{DM}}h^2(\textrm{high }T_R)$ in the standard high
  $T_R$ case for several values of $m_\chi/T_R$.  }
\label{F13}
\end{figure}

Although in practice eq.~(\ref{approxOh2mDM}) is very useful for
understanding the $T_R$-dependence of $\Omega_{\textrm{DM}}h^2$, it
may also be slightly misleading, as it does not show a certain degree of
correlation between $T_\mathrm{fo}$ and
$\Omega_{\textrm{DM}}h^2(\textrm{high }T_R)$. This correlation is easy
to understand, since a large $\Omega_{\textrm{DM}}h^2(\textrm{high
}T_R)$ results from a low (co)annihilation cross-section
which, according to eqs.~(\ref{xFhighTR}) and~(\ref{xFlowTR}), drives
$T_\mathrm{fo}$ to higher values.  An account of this effect is shown
in Figure~\ref{F13}, which shows the relation between
$\Omega_{\textrm{DM}}h^2(\textrm{high }T_R)$ and the true relic
density $\Omega_{\textrm{DM}}h^2$ at some low $T_R$ for different
values of $m_\chi/T_R$.  Obviously, in the high $T_R$ limit
$\Omega_{\textrm{DM}}h^2$ approaches
$\Omega_{\textrm{DM}}h^2(\textrm{high }T_R)$, while for values of
$m_\chi/T_R$ of 20 and more we observe a stronger $T_\mathrm{fo}$
dependence, as predicted by (\ref{approxOh2mDM}), which results in a
slower increase of $\Omega_{\textrm{DM}}h^2$ with growing
$\Omega_{\textrm{DM}}h^2(\textrm{high }T_R)$ and fixed
$m_\chi/T_R$. Of course, if the LOSP is the DM candidate, the
phenomenologically relevant values of $\Omega_{\textrm{DM}}h^2$ belong
to a narrow observed range. However, we shall see in Section
\ref{sec:grav} that for gravitino DM produced in LOSP decays even
larger values of the LOSP relic density will become allowed.

\section{Neutralino dark matter with low reheating temperatures}
\label{sec:neudm}

We will now apply the formalism presented in Section \ref{sec:fo} to
the MSSM with ten free parameters, to the CMSSM, and to the
Next-to-Minimal Supersymmetric Standard Model (NMSSM) with a
singlino-dominated DM.

\subsection{The MSSM\label{secp10mssm}}

In this subsection we will analyze the scenario with low reheating
temperatures of the Universe in the context of the MSSM.  Since a
study of a completely general MSSM would be unmanageable, nor for that
matter even necessary, we select a 10-parameter subset of the MSSM
(p10MSSM) which exhibits all the features of the general model which
are relevant for our discussion.  The free parameters of the model and
their ranges are given in Table~\ref{tabp10MSSM}. Our choice follows
that of \cite{Fowlie:2013} (see discussion therein), except that we
keep both the wino mass $M_2$ and the bino mass
$M_1$ free in order to allow each of them to be DM. As we will see,
the choice of ten free parameters will allow various accidental
mass degeneracies which can contribute to coannihilations.  Also, the
ranges of parameters have been extended to obtain a wide range of
$\Omega_{\textrm{DM}}h^2 (\textrm{high }T_R)$ with $m_{\textrm{DM}}$
reaching up to $5$ TeV.

%%%%%%%%%%%%%%%%%%
\begin{table}[h]
\centering
\begin{tabular}{|c|c|}
\hline 
Parameter & Range \\ 
\hline
\hline 
bino mass & $0.1 < M_1 < 5$ \\ 
wino mass & $0.1 < M_2 < 6$ \\ 
gluino mass & $0.7 < M_3 < 10$ \\ 
stop trilinear coupling & $-12 < A_t < 12$ \\ 
stau trilinear coupling & $-12 < A_{\tau} < 12$ \\ 
sbottom trilinear coupling & $A_b = -0.5$ \\ 
pseudoscalar mass & $0.2 < m_A < 10$ \\ 
$\mu$ parameter & $0.1 < \mu < 6$ \\ 
3rd gen. soft squark mass & $0.1 < m_{\widetilde{Q}_3} < 15$ \\ 
3rd gen. soft slepton mass & $0.1 < m_{\widetilde{L}_3} < 15$ \\ 
1st/2nd gen. soft squark mass  & $m_{\widetilde{Q}_{1,2}} = M_1 + 100$ GeV \\ 
1st/2nd gen. soft slepton mass  & $m_{\widetilde{L}_{1,2}} = m_{\widetilde{Q}_3} + 1$ TeV \\ 
ratio of Higgs doublet VEVs & $2 < \tan\beta < 62$ \\ 
\hline 
\hline
Nuisance parameter & Central value, error \\
\hline
\hline
Bottom mass \mbmbmsbar (GeV) & (4.18, 0.03) \cite{PDG:2014}\\
Top pole mass \mtop (GeV) & (173.5, 1.0) \cite{PDG:2014}\\
\hline
\end{tabular}
\caption{The parameters of the p10MSSM and their
  ranges used in our scan. All masses and trilinear couplings are given in TeV, unless
  indicated otherwise. All the parameters of the model are given at the
  SUSY breaking scale.} 
\label{tabp10MSSM} 
\end{table}
%%%%%%%%%%%%%%%%%%

We scan the parameter space of p10MSSM following the Bayesian
approach. The numerical analysis was performed using the BayesFITS package
which engages Multinest \cite{Feroz:2009} for sampling the parameter
space of the model. Supersymmetric mass spectra were calculated with
SOFTSUSY-3.4.0 \cite{Allanach:2002}, while $B$-physics related
quantities with SuperIso v3.3 \cite{Arbey:2009}. MicrOMEGAs v3.6.7
\cite{Belanger:2013} was used to obtain
$\Omega_{\textrm{DM}}h^2(\textrm{high }T_R)$ and DM-proton
spin-independent direct detection cross section
$\sigma_{p}^{\textrm{SI}}$.

%%%%%%%%%%%%%%%%%%
\begin{table}[t]
\centering
\begin{tabular}{|c|c|c|c|}
\hline 
Measurement & Mean & Error: exp., theor. & Ref. \\ 
\hline
\hline 
$m_h$ & $125.7$ GeV  & $0.4$ GeV, $3$ GeV & \cite{CMS:2013}\\ 
\abunchi & $0.1199$ & $0.0027$, $10\%$ & \cite{Planck:2013}\\ 
\brbxsgamma$\times 10^4$ & $3.43$ & $0.22$, $0.21$ & \cite{SLAC} \\ 
\brbutaunu$\times 10^4$ & $0.72$ & $0.27$, $0.38$ & \cite{Belle:2012}\\ 
\delmbs & $17.719$ $\textrm{ps}^{-1}$ & $0.043\,\textrm{ps}^{-1}$, $2.400\,\textrm{ps}^{-1}$ & \cite{PDG:2014}\\
\sinsqeff & $0.23116$& $0.00013$, $0.00015$ & \cite{PDG:2014}\\
$M_W$ & $80.385$ GeV & $0.015$ GeV, $0.015$ GeV & \cite{PDG:2014}\\
\brbsmumu$\times 10^9$ & $2.9$ & $0.7$, $10\%$ & \cite{LHCb:2013,CMS:2013Jul}\\
\hline 
\end{tabular}
\caption{The constraints imposed on the parameter spaces of the p10MSSM and
the  CMSSM. The LUX upper limits \cite{LUX:2013} have been implemented as a hard cut.}
\label{constraints} 
\end{table}
%%%%%%%%%%%%%%%%%%

The constraints imposed  in scans are listed in
Table~\ref{constraints}. The LHC limits for supersymmetric particle
masses were implemented following the methodology described in
\cite{Fowlie:2013,Fowlie:2012}.  The DM relic density for low $T_R$
was calculated by solving numerically the set of Boltzmann equations
(\ref{Boltzmann}), as outlined in \cite{Giudice:2000}. In order to
find the point where WIMPs freeze-out we adapted the method described,
e.g., in \cite{Bhupal:2013} to the scenario with a low reheating
temperature. Both $\langle\sigma v\rangle_{\textrm{eff}}$ and
$\langle\sigma v\rangle_{\textrm{eff}}\langle E\rangle_{\textrm{eff}}$
as a function of temperature were obtained with appropriately modified
MicrOMEGAs; we also checked that in the high $T_R$ limit we reproduced
$\Omega_{\textrm{DM}}h^2$ obtained with the original version of this
code.

\begin{figure}[t]
\begin{center}
\includegraphics*[width=9.5cm,height=8.5cm]{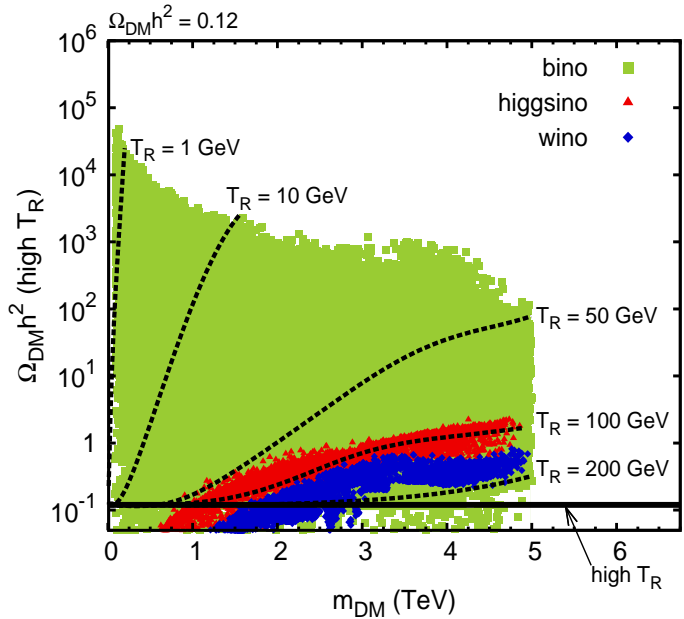}
\end{center}
\caption{Contours (black dotted) of constant $\Omega_{\textrm{DM}}h^2 = 0.12$ for different values of the reheating
  temperature $T_R$ in the MSSM in the $(m_{\rm{DM}},
  \Omega_{\rm{DM}}h^2(\textrm{high }T_R)\,)$ plane. The solid black
  horizontal line corresponds to the high $T_R$ limit. Green squares
  correspond to the bino DM region, while red triangles (blue
  diamonds) to the higgsino (wino) DM case. }
\label{MSSMclueTR}
\end{figure}

The results of the scans -- but without imposing the constraint on the
DM relic abundance and direct detection rates -- are shown in
Figure~\ref{MSSMclueTR}, with lines of constant $T_R = 1, 10, 50, 100,
200$ GeV superimposed along which $\Omega_{\textrm{DM}}h^2 \simeq 0.12$. The horizontal line corresponds to the correct DM
relic density in the standard high $T_R$ scenario. Different colours
denotes various compositions of the lightest neutralino: 
green, red and blue corresponds to the bino, higgsino and wino fraction
larger than 95\%.
We will now describe the results for each of
these three cases.

%%%%%%%%%%%%%%%%%%
%%%%%%%%%%%%%%%%%%

%%%%%%%%%%%%%%%%%%
\paragraph*{Bino DM.}
The region of bino DM covers most of the plane in
Figure~\ref{MSSMclueTR}.  In this case the relic density can vary by
several orders of magnitude for a given $m_{\textrm{DM}}$, since it
is very sensitive to the details of the MSSM spectrum.  Generically,
bino annihilation rate is dominated by $t$-channel slepton exchange
$\chi\chi \rightarrow l\bar{l}$ and for $m_{\widetilde{B}}\ll
m_{\widetilde{l}}$ the bino relic density reads (see,
e.g.,  \cite{Drees:1992,Wells:1998})
%%%%%%%%%%%%%%%%%%
\begin{equation}
\Omega_{\widetilde{B}}h^2(\textrm{high }T_R) \approx 
%g_{*,\mathrm{fo}}^{-1/2} \frac{1}{M^2\,\sqrt{g^{\ast}_{\rm{fo}}}}\,\frac{(m_{\widetilde{l}}^2 + m_{\widetilde{B}}^2)^4}{m_{\widetilde{B}}^2\,(m_{\widetilde{l}}^4 + m_{\widetilde{B}}^4)} \xrightarrow{m_{\widetilde{l}}\gg m_{\widetilde{B}}} 
g_{\ast,\mathrm{fo}}^{-1/2}
\,\left(\frac{m_{\widetilde{l}}}{m_{\widetilde{B}}}\right)^2
\,\left(\frac{m_{\widetilde{l}}}{460\,\mathrm{GeV}}\right)^2 \, ,
\label{eq:ombino}
\end{equation}
%%%%%%%%%%%%%%%%%%
where $g^{\ast}_{\rm{fo}}$ stands for the number of relativistic
degrees of freedom at $\chi$ decoupling.  By varying the bino and the
slepton masses, one can obtain $\Omega_{\widetilde{B}}h^2(\textrm{high
}T_R)$ spanning a few orders of magnitude. The upper boundary of
the allowed region in Figure~\ref{MSSMclueTR} has no physical meaning --
it simply corresponds to the maximum value of slepton masses in our
scan which is $\sim 10 - 15$~TeV.

It is well-known that the correct
$\Omega_{\widetilde{B}}h^2(\textrm{high }T_R)$ can be achieved for low
$m_{\widetilde{B}}$ typically thanks to coannihilations with the
lighter stau or, for $m_A \simeq 2m_{\widetilde{B}}$, to resonant
annihilations through the $s$-channel exchange of the pseudoscalar Higgs
boson ($A$-funnel region); however,
$\Omega_{\widetilde{B}}h^2(\textrm{high }T_R)\sim 0.12$ can also be
obtained with the lighter Higgs boson $h$ resonance \cite{Ellis:1989pg}, for bino-higgsino
mixing or in the bulk region where the bino-dominated neutralino
annihilates through $t$-channel exchange of sfermions (typically of
sleptons as they are usually lighter than squarks -- see a discussion of these regions
in, e.g., \cite{Fowlie:2013}).  Note that, for large bino mass, $m_{\widetilde{B}}
> 2$ TeV, its relic density can still be reduced by the Higgs
pseudoscalar exchange in the $A$-funnel region, but also through
coannihilations owing to accidental 
bino-wino or bino-gluino mass degeneracies.  This explains the presence
of points with a very low bino relic density at large WIMP mass in Figure~\ref{MSSMclueTR}.

\paragraph*{Higgsino DM.}
The results for the higgsino DM relic density agree well with other
recent analyses (see, e.g., \cite{Fowlie:2013}). In
Figure~\ref{MSSMclueTR}, $\Omega_{\textrm{DM}}h^2(\textrm{high }T_R)$
scales proportionally to $m_{\textrm{DM}}^2$, achieving the correct
value at $m_{\textrm{DM}}\sim 1$ TeV. However, one can see that,
for the whole range above that value one can obtain the observed value
of the relic density provided $T_R$ is low enough, around 100~GeV.

\paragraph*{Wino DM.}
Wino relic density is quite sensitive to a so-called Sommerfeld
enhancement (SE) of the annihilation cross-section due to attractive
Yukawa potentials induced by the electroweak gauge bosons
\cite{Hisano:2003ec} (see also, e.g., \cite{Hryczuk:2010} for a recent
and exhaustive discussion; we use enhancement factors from that
reference in our numerical analysis). Incidentally, the SE is particularly
important in the $\sim 2- 3$ TeV wino mass range, where the correct
$\Omega_{\widetilde{W}}h^2$ can be obtained for high $T_R$.  In our
scan, the SE is responsible for a visible vertical broadening of the wino region around
$~2.5$~TeV.

When considering the wino as a DM candidate, one has to take into
account that the SE is associated with enhanced rates of present-day
wino annihilations giving rise to diffuse gamma ray background;
therefore, stringent indirect detection bounds apply in this case.  It
has been shown \cite{Cohen:2013,Fan:2013,Hryczuk:2014} that the
enhancement of indirect detection rates for $m_{\widetilde{W}}
\lesssim 3.5$ TeV is in conflict with current observational limits. On
the other hand, wino DM with mass larger than 3.5~TeV generically
has too large relic abundance, which excludes it as a DM candidate
over the whole mass range in the standard high $T_R$ scenario.

\paragraph*{}
For each of the three neutralino compositions discussed above, a
suppression of the DM relic abundance at low $T_R$ leads to
interesting, and often dramatic, consequences, allowing vast regions
of the parameters space regarded as phenomenologically disallowed in
the high $T_R$ limit. In the following we shall present a more
detailed analysis of the parameter space of the MSSM with low $T_R$.

Scenarios with a low reheating temperature allow choices of the MSSM
parameters which at high $T_R$   would lead to too small DM annihilation rates and,
as a consequence, too large relic density.  Since
small annihilation rates are usually associated with small direct
detection rates, it is interesting to analyze the
effect of the assumed low reheating temperature.  We shall 
discuss here both the most recent constraints from the LUX experiment
\cite{LUX:2013}, as well as from expected future reach of the one-tonne
extension of the Xenon experiment (Xenon1T) \cite{Aprile:2012zx}.

%%%%%%%%%%%%%%
\begin{figure}[!t]
\begin{center}
\includegraphics*[width=8cm,height=7cm]{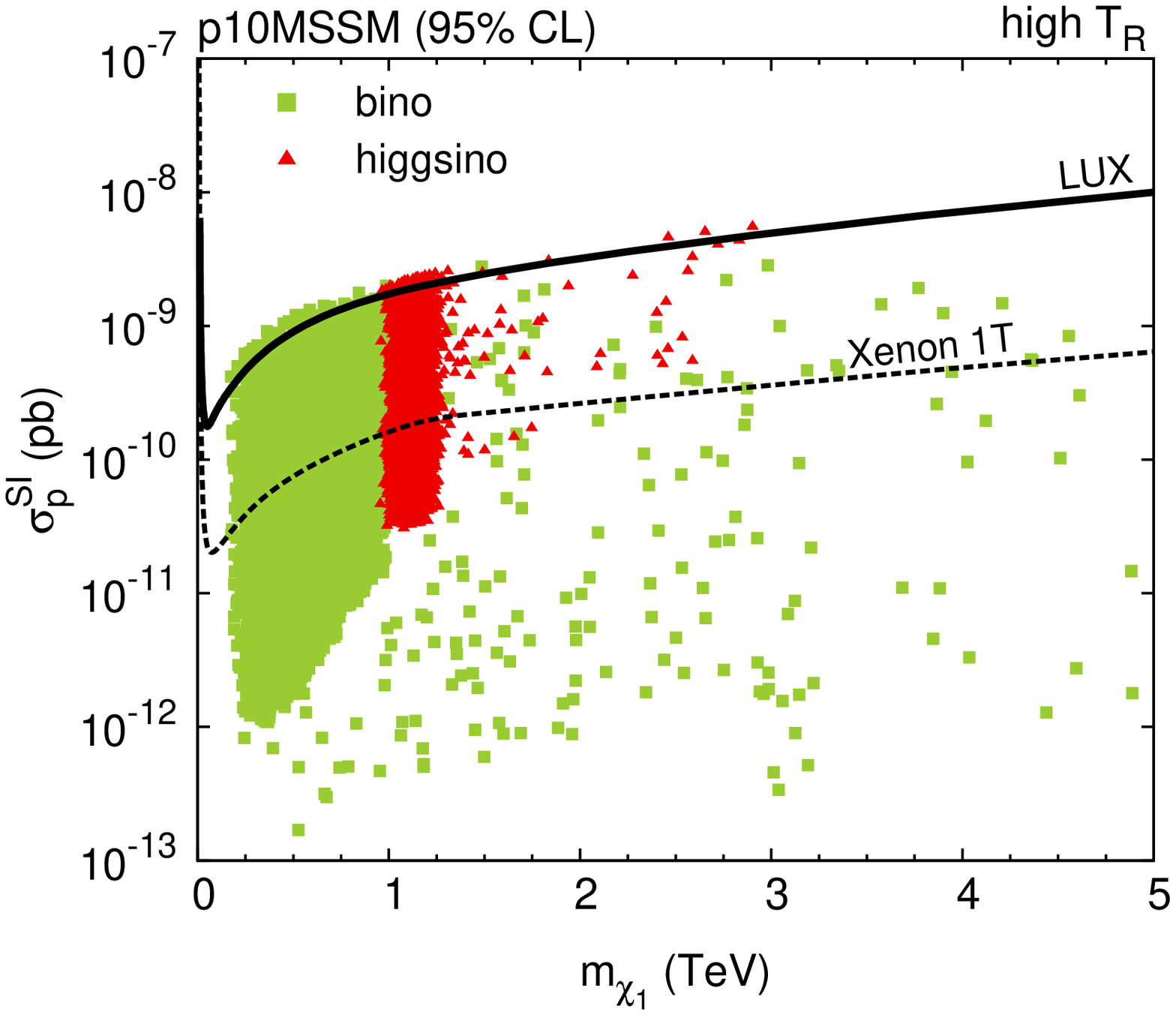}
\hfill
\includegraphics*[width=8cm,height=7cm]{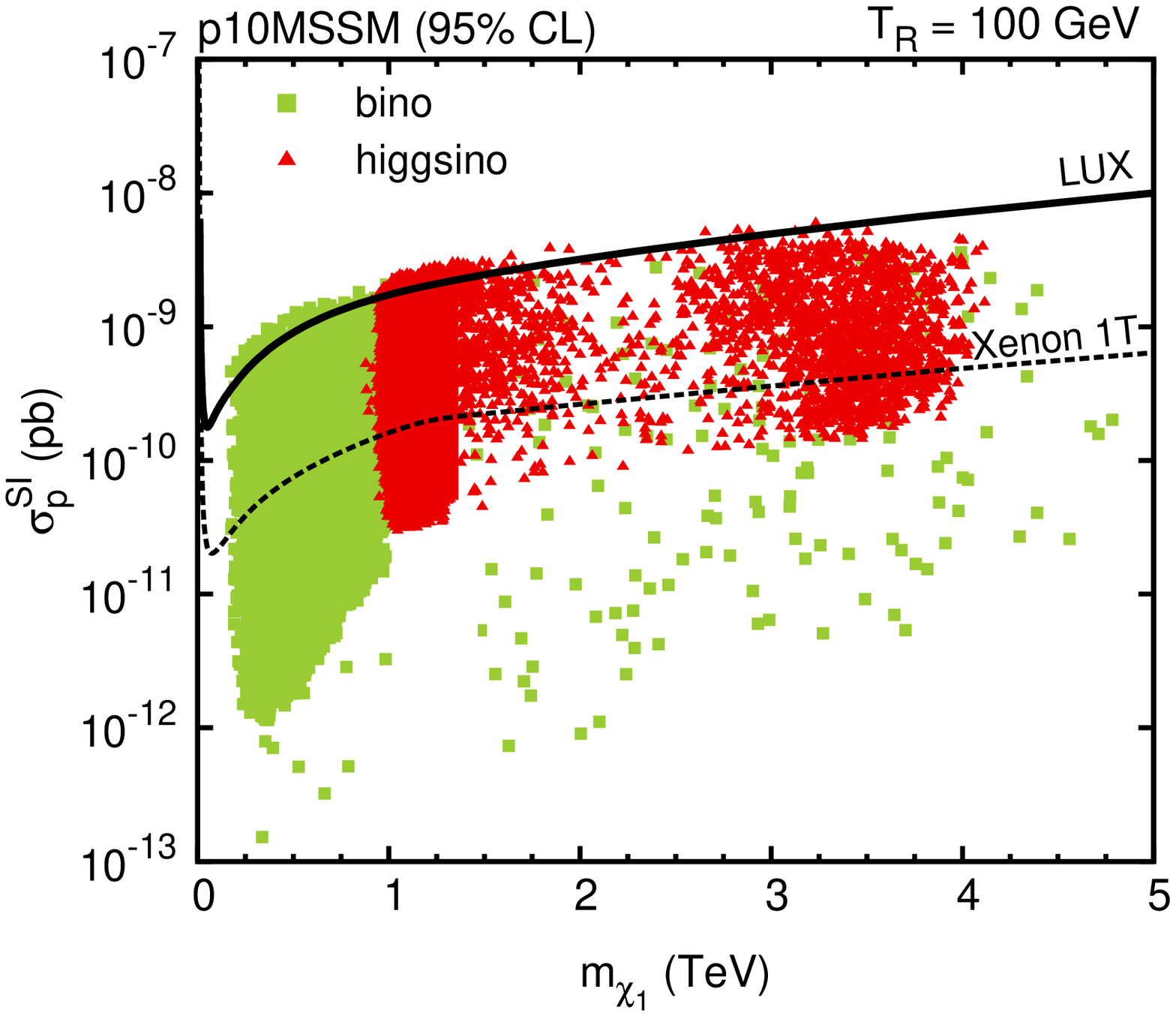}

\includegraphics*[width=8cm,height=7cm]{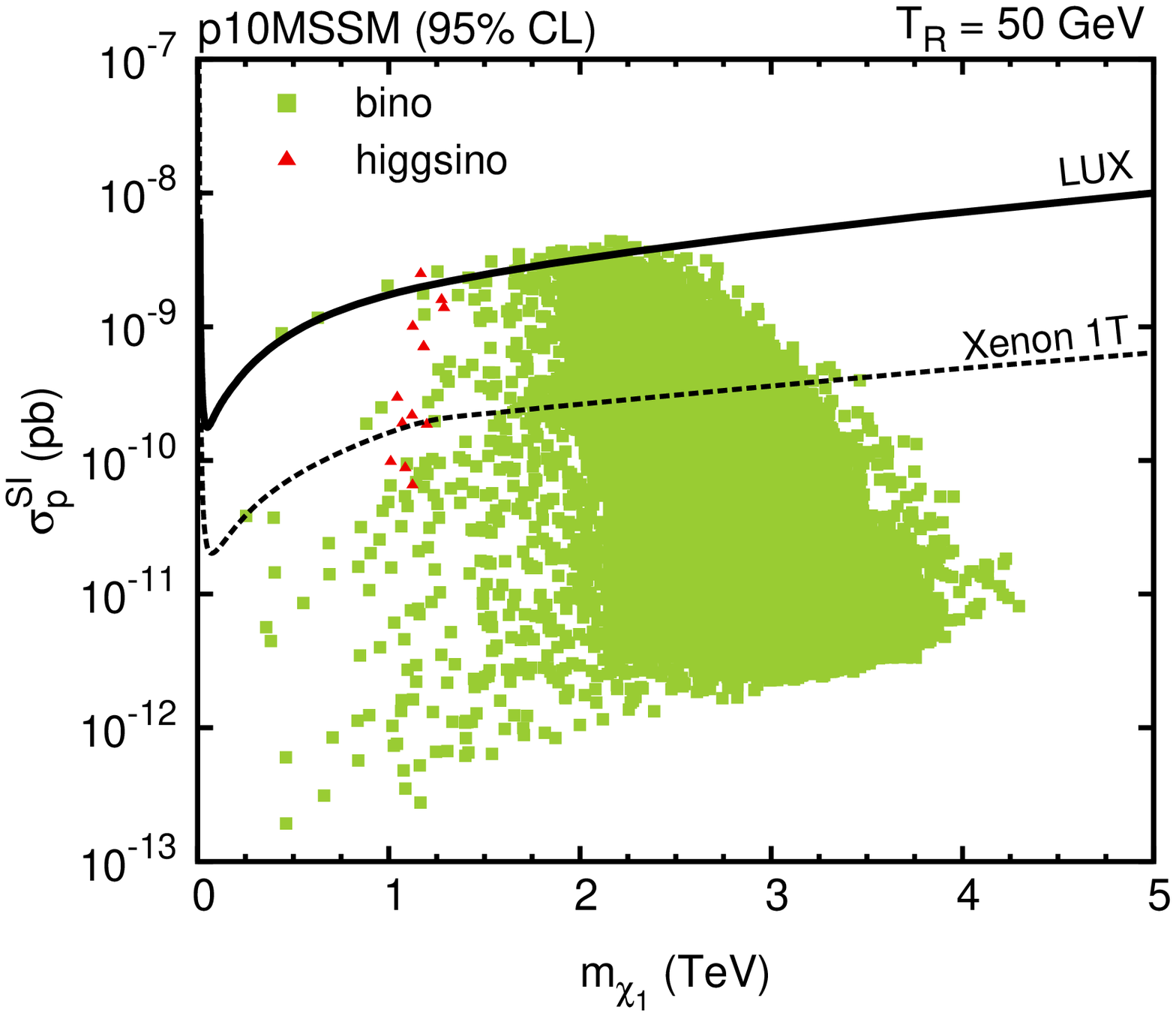}
\hfill
\includegraphics*[width=8cm,height=7cm]{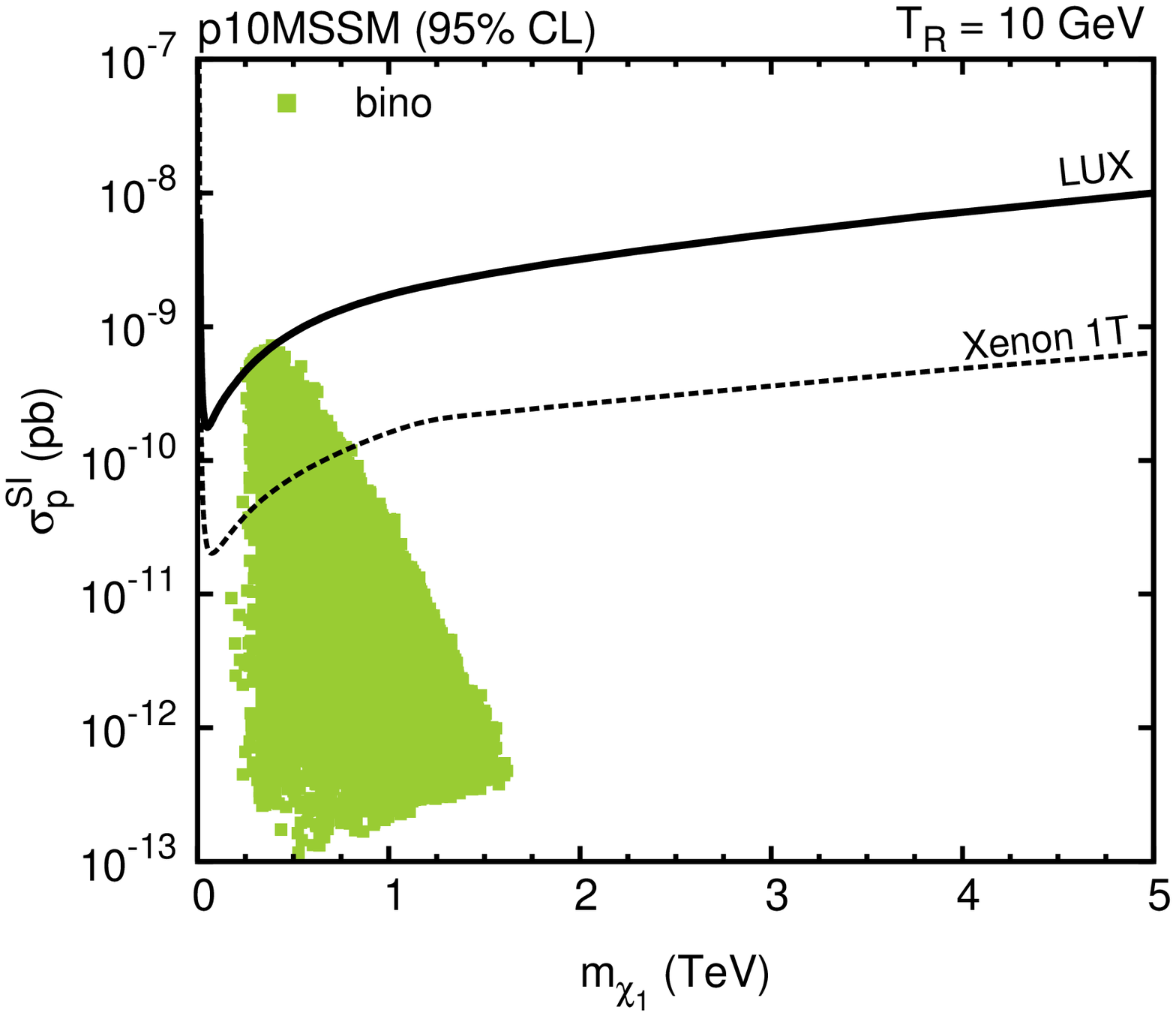}
\end{center}
\caption{Direct detection \sigsip\ cross section as a function of
  $m_{\chi_1}$ in the p10MSSM $2\sigma$ credible regions for several
  fixed values of the reheating temperature. The solid (dashed) black
  lines correspond to LUX (projected Xenon1T) limit on \sigsip. 
  Color coding as in Figure~\ref{MSSMclueTR}.
}
\label{p10MSSMsigsip}
\end{figure}
%%%%%%%%%%%%%%

In Figure~\ref{p10MSSMsigsip} we show -- for fixed values of $T_R$ --
the $2\sigma$ credible regions in the $(m_{\chi},\sigsip)$ plane for
the p10MSSM scans with the DM density constraint included. In the case of high
reheating temperature (upper left panel) most points correspond to $m_{\chi} \lesssim
1.5$~TeV: these are either bino- or higgsino-like neutralinos.  
 Scenarios in which the neutralino is the bino with a few per cent higgsino
admixture
are typically characterised by enhanced
\sigsip; such points occupy the upper part of the bino DM (green)
region and will be accessible to Xenon1T. An
almost pure bino neutralino, instead, can have much lower direct
detection cross-section and it often remains beyond the reach of
current and future experiments. In the case of higgsino DM, a good
fraction of points lie within the projected Xenon1T sensitivity.  As
we have discussed in Section \ref{secp10mssm}, for higher $m_{\chi}$ one
needs specific mass patterns to obtain the correct relic density; as
these are fine-tuned cases, one obtains fewer points for $m_{\chi}
\gtrsim 1.5$ TeV than for lower DM mass values. The wino, which can
have the correct relic density for $m_{\widetilde{W}} \sim 2-3$ TeV,
is not shown in the plot, since it is excluded by the indirect DM
searches in this mass range \cite{Cohen:2013,Fan:2013,Hryczuk:2014}.
%%%%%%%%%%%%%%
\begin{figure}[!t]
\begin{center}
\includegraphics*[width=8cm,height=7cm]{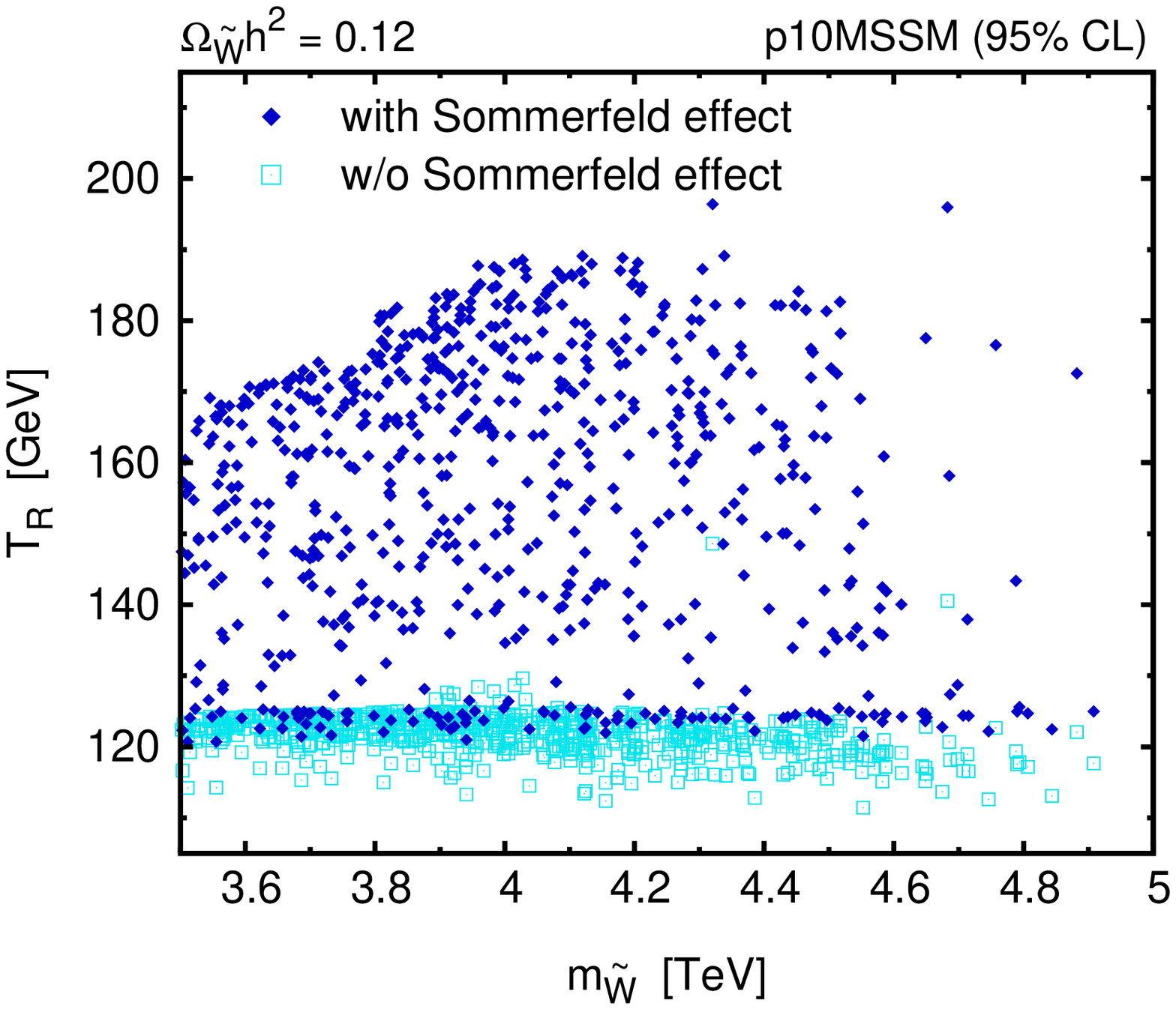}
\hfill
\includegraphics*[width=8cm,height=7cm]{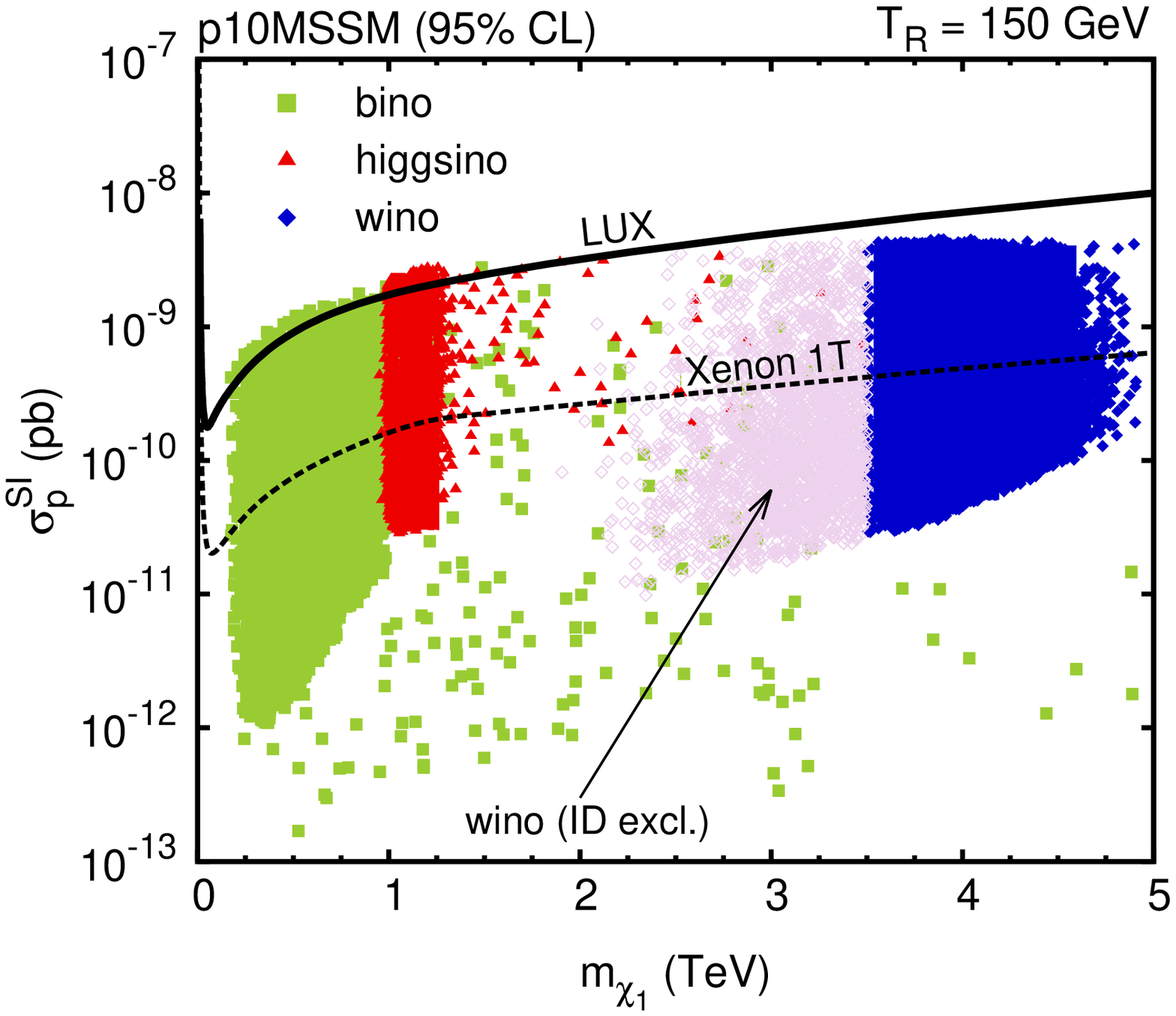}
\end{center}
\caption{{\em Left panel}: the reheating temperature range in the wino
  DM scenario that gives the correct relic density for
  $m_{\widetilde{W}} > 3.5$ TeV where indirect detection limits are
  not violated. The results with (without) the Sommerfeld effect are
  shown as dark blue solid diamonds (light blue empty squares). {\em
    Right panel}: the $2\sigma$ credible region of the p10MSSM for
  $T_R = 150$ GeV in the $(m_{\chi},\sigsip)$ plane with the
  Sommerfeld effect included in calculating the relic density. In the
  case of wino DM, we use pink (blue) color to distinguish points
  which are excluded (not excluded) by the requirement
  $m_{\widetilde{W}} > 3.5$~TeV imposed by indirect detection
  searches.  The solid (dashed) black line corresponds to the LUX (a
  projected Xenon1T) limit on \sigsip. Remaining color coding as in
  Figure~\ref{MSSMclueTR}.}
\label{wino}
\end{figure}
%%%%%%%%%%%%%%

As expected from Figure~\ref{MSSMclueTR}, for $T_R = 100$ GeV  (upper
right panel) the
results in the low $m_{\chi}$ region are virtually the same as for
high $T_R$.  However, an important difference appears at $m_{\chi}
\sim 3 - 4$ TeV where one can obtain the desired value
$\Omega_{\chi}h^2 \simeq 0.12$ for the higgsino. In
this  region, the direct detection cross section \sigsip\ is
high enough to allow testing the scenario by the Xenon1T experiment.
We also note that, though Figure~\ref{MSSMclueTR} suggests that for $T_R
= 100$~GeV one can have a higgsino-like DM with any mass in the
scanned range, higgsino mass between $2\
\textrm{TeV}\,\,\textrm{and}\,\, 2.5$~TeV are disfavored because of
$\Omega_{\chi}h^2$ being often too large.
 As a result one
observes a reduced number of higgsino-like points in this mass range.

For $T_R = 50$ GeV  (lower left panel), the low-$T_R$ relic density suppression is already
effective for $m_\chi\sim1$~TeV and it is very strong for larger DM
mass, making the higgsino strongly disfavoured.  We can see just a few $\sim1$~TeV higgsino-like
neutralinos characterised by $\Omega_{\chi}h^2(\textrm{high }T_R) \sim
0.2 - 0.4$.  On the other hand, for $m_\chi>1$~TeV one can now easily
obtain the correct relic density for a nearly pure bino without
requiring any specific relation among soft supersymmetry (SUSY) breaking
parameters.  The region with $m_{\chi} < 1$ TeV now becomes less
appealing, since it still requires some specific mass pattern to
suppress the relic density, and we find only a few points there. As
can be seen in Figure~\ref{p10MSSMsigsip}, only a fraction of the $2\sigma$
credible region lies above the Xenon1T expected reach in the range of $\sim
2-3$ TeV mass.

For $T_R = 10$ GeV (lower right panel), only points corresponding to $m_{\chi} <
1.5$ TeV are present in our scan.  This feature does not have a
physical origin, but it merely results from a finite, albeit generous,
ranges of the superpartner masses which we have allowed; this limit
can be seen in Figure~\ref{MSSMclueTR}.  Since low-$T_R$ suppression is
now very effective in the entire DM mass range, these points typically
have large $\Omega_\mathrm{DM}h^2(\textrm{high }T_R)$, hence low
\sigsip\ and the experimental verification of such scenarios poses a
challenge.

With the values of $T_R$ discussed so far we have not seen any
acceptable points corresponding to wino DM. This can be easily understood by
examining Figure~\ref{MSSMclueTR} which shows that the wino DM with
$m_\chi\simgt 3.5$~TeV has the correct relic density for $T_R$ only
between 100~and 200~GeV.  The left panel of Figure~\ref{wino} shows the
reheating temperature for the points in the $2\sigma$ credible region
in the p10MSSM corresponding to the wino with the correct
abundance -- with and without the SE taken into account.  With the SE
neglected, the points form a narrow band with $T_R$ between 115 and
120~GeV.  Since the SE leads to a suppression of
$\Omega_{\chi}h^2(\textrm{high }T_R)$, its inclusion allows one to
obtain the measured DM relic density for slightly larger $T_R$. The
actual enhancement of the cross-section depends on the value of $\mu$
and can therefore vary for a given wino mass. Hence,
including the SE one obtains $\Omega_{\widetilde{W}}h^2
\simeq 0.12$ for a wider range of reheating temperatures
$120\,\textrm{GeV}\lesssim T_R \lesssim 200$ GeV.  In the right panel
of Figure~\ref{wino} we show -- for $T_R = 150$~GeV -- the $2\sigma$
credible region of the p10MSSM on $(m_{\chi}, \sigsip)$ plane. Regions
with lower $m_{\chi}$ corresponding to the bino or the higgsino are
similar to the high $T_R$ case as expected. At larger $m_\chi$, a new
region with the wino DM becomes allowed for mass of $\gtrsim 3.5$
TeV. It is not excluded by current limits from indirect detection
experiments, but potentially can be in the future
\cite{Hryczuk:2014}. Although some of these points lie within the
projected Xenon1T sensitivity reach, direct detection experiments will not
constrain the scenario too strongly.

One may wonder whether the lighter Higgs boson mass, $m_h \approx 126$
GeV, constrains the low-$T_R$ bino DM scenarios in any significant
way. The answer is negative:
one obtains a sufficiently large $m_h$ by arranging
large stop masses and/or a large left-right mixing in the stop sector, while the bino relic density depends mainly on bino and stau masses.
Since in 
the p10MSSM  discussed in this section 
the stop and bino/stau sectors are to a large degree independent,
for all the points presented in Figure~\ref{p10MSSMsigsip} the lighter
Higgs boson mass comes out close to the experimentally measured value
thanks to heavy squarks, well above the LHC
limits for colored superpartners.

\subsection{The CMSSM}

%%%%%%%%%%%%%%
\begin{table}[t]
\centering
\begin{tabular}{|c|c|}
\hline 
Parameter & Range \\ 
\hline
\hline 
common scalar mass & $0.1 < m_0 < 10$ \\ 
common gaugino mass & $0.1 < m_{1/2} < 10$ \\ 
common trilinear coupling & $-15 < A_0 < 15$ \\ 
ratio of Higgs doublet VEVs & $2 < \tan\beta < 62$ \\ 
sign of $\mu$ parameter & $\mu > 0$\\
\hline 
\end{tabular}
\caption{The parameters of the CMSSM and their ranges used in our scan. All
  masses and trilinear couplings are given in TeV, unless indicated
  otherwise. Masses and trilinear coupling are given at the GUT
  scale. The nuisance parameters are the same as for the p10MSSM.} 
\label{tabCMSSM} 
\end{table}
%%%%%%%%%%%%%%

%%%%%%%%%%%%%%
\begin{figure}[!t]
\begin{center}
\includegraphics*[width=8cm,height=7cm]{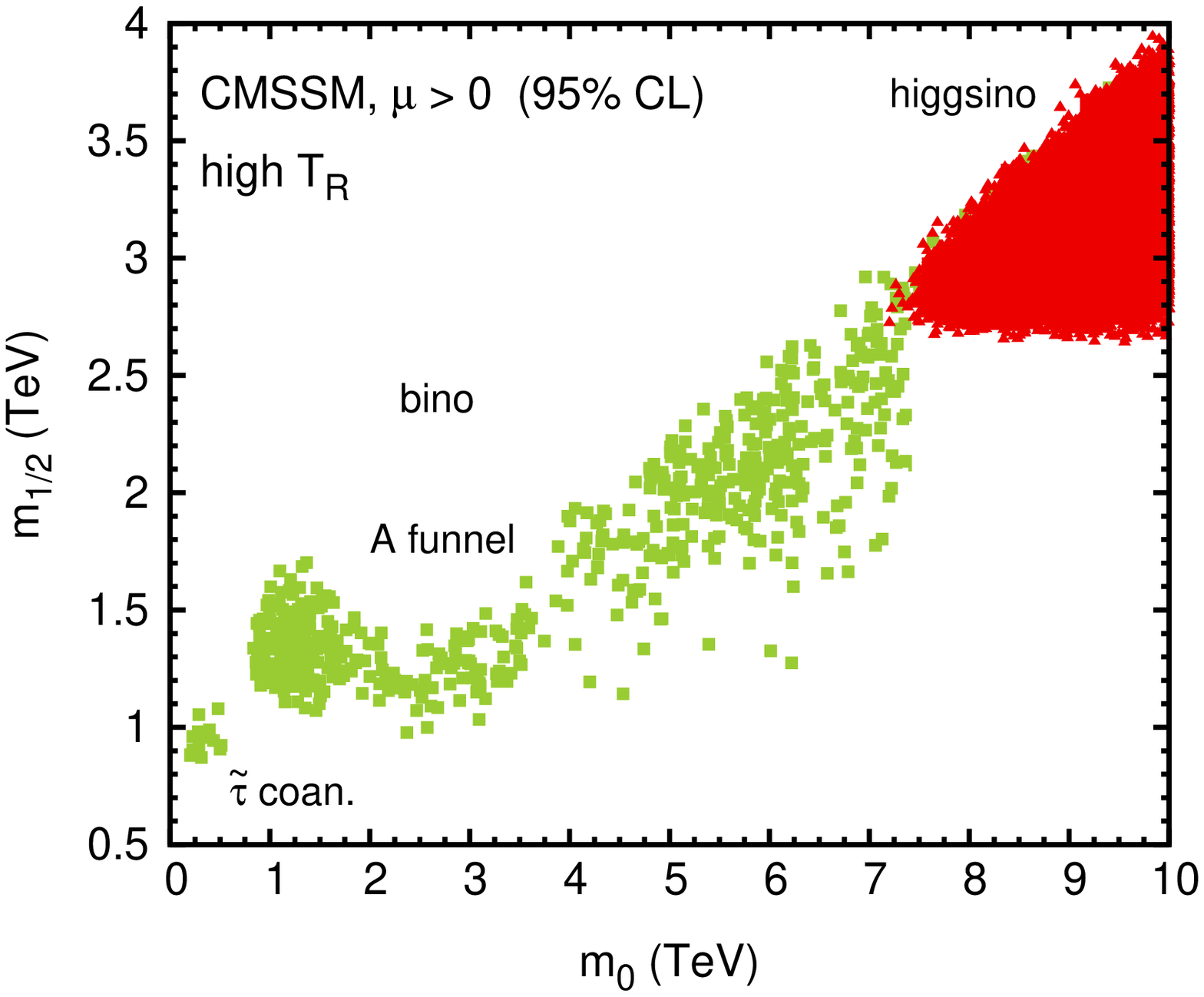}
\hfill
\includegraphics*[width=8cm,height=7cm]{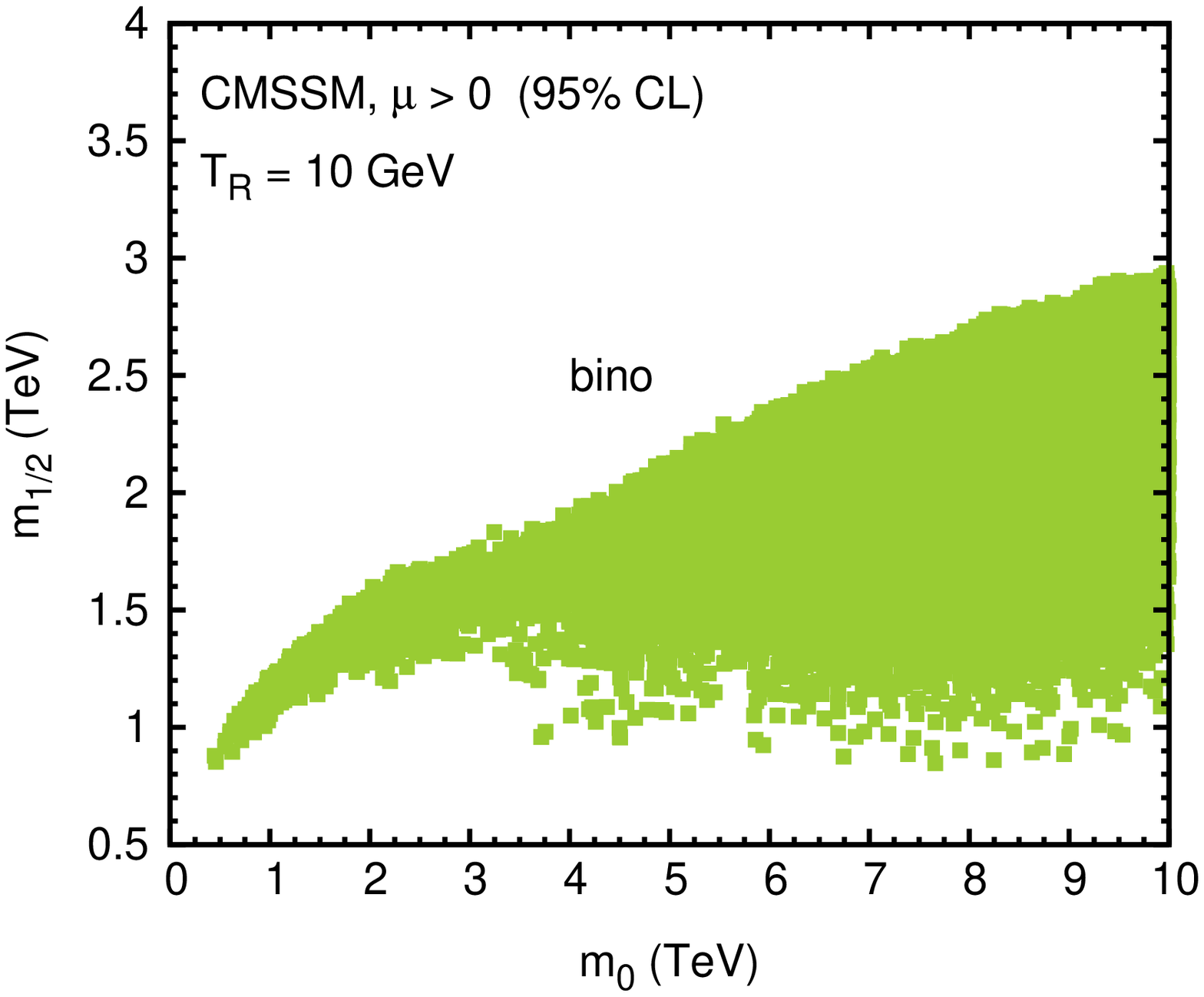}
\end{center}
\caption{The $2\sigma$ credible regions in the $(m_0,m_{1/2})$ plane of
  the CMSSM for high reheating temperature (left panel) and $T_R = 10$
  GeV (right panel).}
\label{CMSSMm0m12}
\end{figure}
%%%%%%%%%%%%%%

We will now examine which features, if any, of the general MSSM with
low reheating temperature will remain when we relate its many free
parameters by the assumption of a grand unification.  A prime example
of this class of models is the Constrained MSSM~\cite{Kane:1993td},
where unification conditions are imposed at the GUT scale.  The
parameters of CMSSM and their ranges are given in
Table~\ref{tabCMSSM}.

In Figure~\ref{CMSSMm0m12} we show $2\sigma$ credible regions of the
($m_0$,$m_{1/2}$) plane with high (left panel) and low $T_R = 10$ GeV
(right panel). In the high-$T_R$ scenario one can identify three
well-known regions with low $\chi^2$ (see, e.g.,~\cite{Fowlie:2012})
that correspond to the correct relic density of neutralino DM: from
left to right, the stau coannihilation and the $A$-funnel regions, as
well as the $\sim
1$ TeV higgsino  region. The focus-point/hyperbolic branch region is
absent, since it has been excluded by the LUX limit on DM direct
detection cross section for positive $\mu$.

%%%%%%%%%%%%%%
\begin{figure}[!t]
\begin{center}
\includegraphics*[width=8cm,height=7cm]{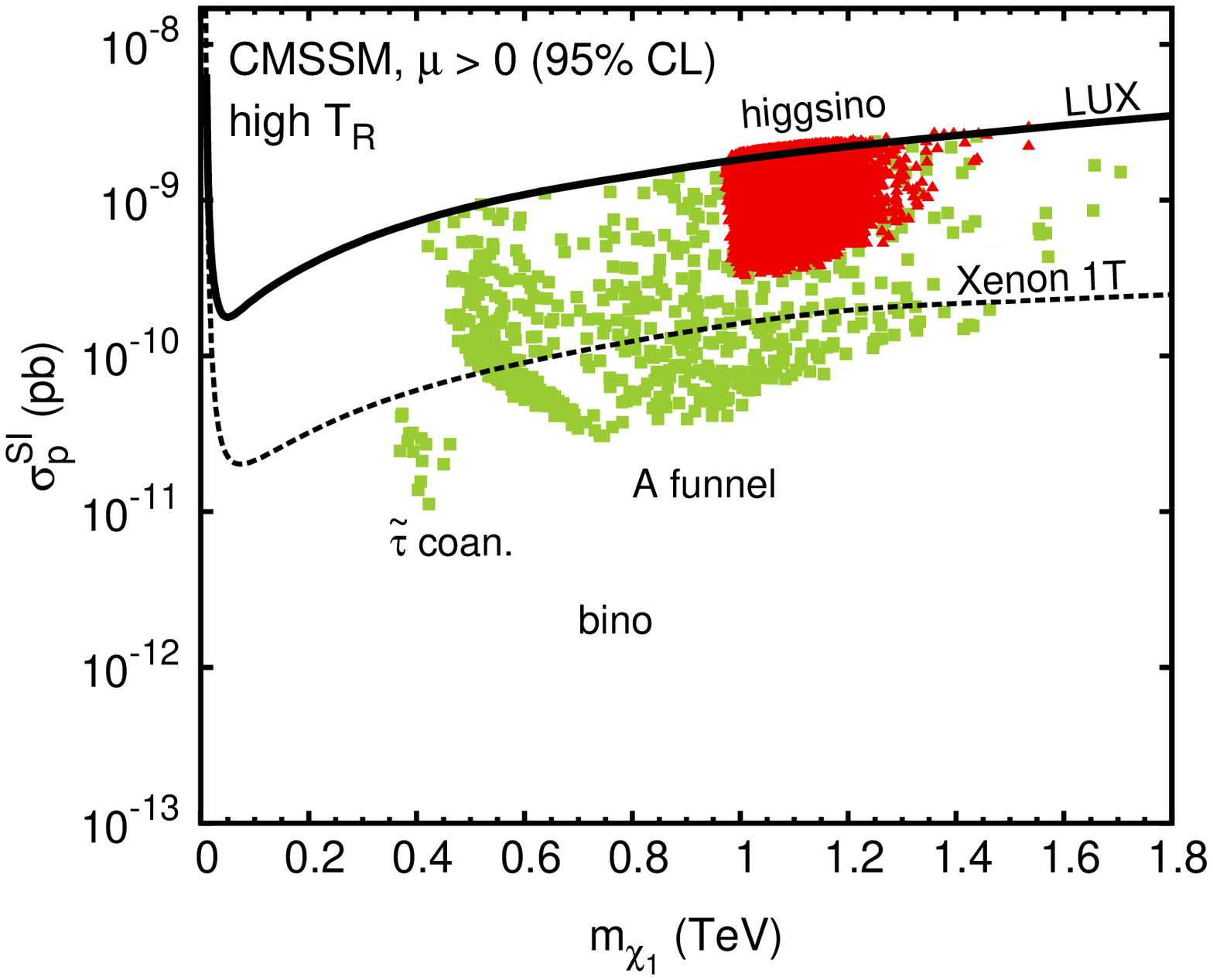}
\hfill
\includegraphics*[width=8cm,height=7cm]{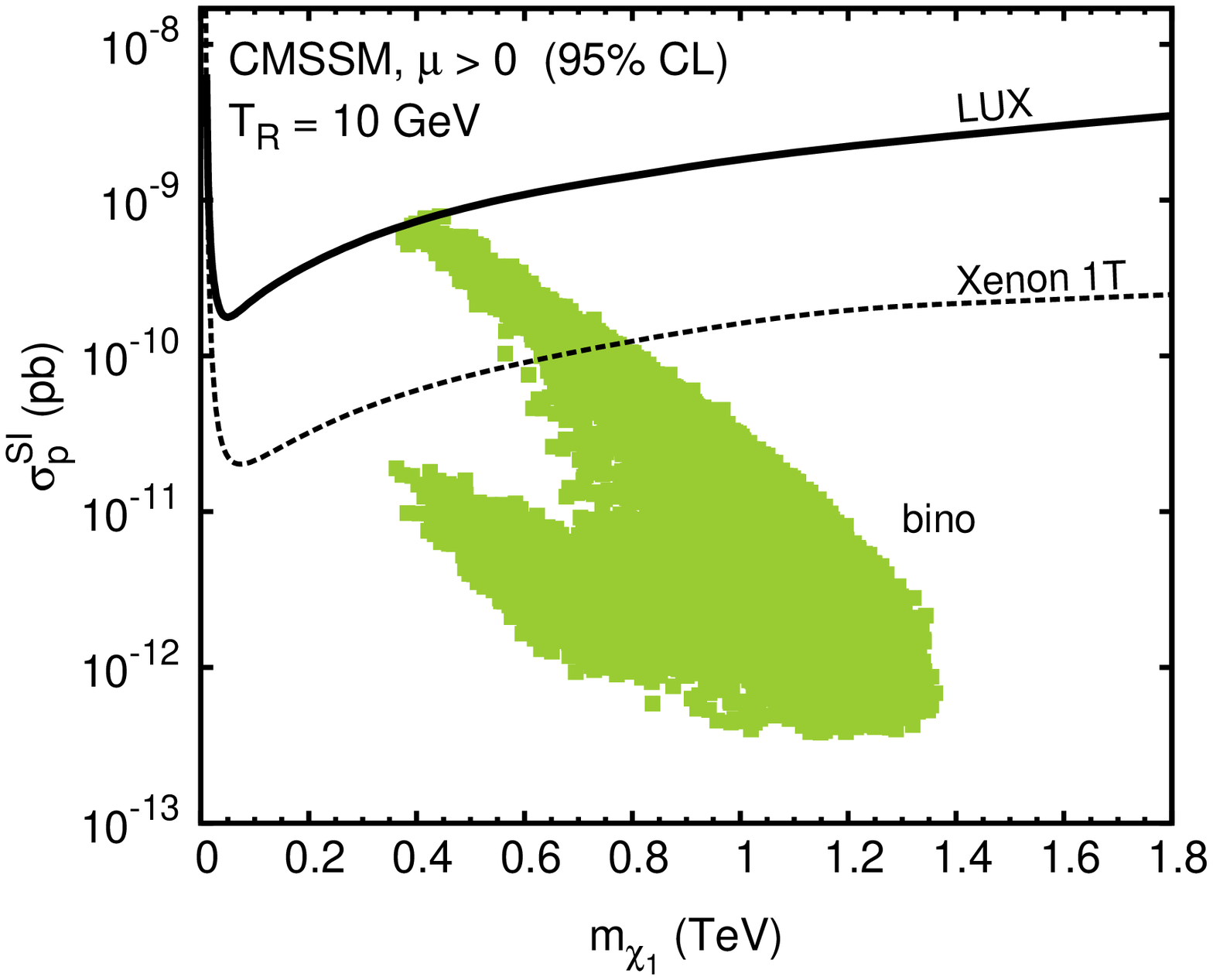}
\end{center}
\caption{The direct detection \sigsip\ cross section as a function of
  $m_{\chi_1}$ for the CMSSM $2\sigma$ credible regions at high
  reheating temperatures (left panel) and for $T_R = 10$ GeV (right
  panel).}
\label{CMSSMsigsip}
\end{figure}

As we have seen in Figure~\ref{MSSMclueTR}, for $T_R = 10$ GeV only the bino
can produce the correct relic density. The lower left corner of
the allowed region in ($m_0$,$m_{1/2}$) plane corresponds to stau
coannihilation region, analogous to that obtained for high $T_R$, (for
such low WIMP mass values the suppression due to low $T_R$ is inefficient).
For slightly higher values of the mass parameters, the suppression of
the relic density by stau coannihilations is traded for low-$T_R$
suppression and we find acceptable points there.  In that region, the bino relic density for a fixed
$T_R$ and a fixed bino mass (or $m_{1/2}$)
depends on many factors, in particular, on stau masses (which depend
not only on $m_0$, but also on $\tan\beta$ and $A_0$), as well as on the small
but non-negligible higgsino fraction of the lightest neutralino.

Unlike in the general MSSM, in both the high- and low-$T_R$ regime the
the Higgs boson mass and the DM relic density depend in
part on the
same parameters of the model, so they are not completely independent. This is
illustrated by the case of $T_R=10$~GeV. It is known that unless 
stop  masses are in the few-TeV range, the Higgs boson mass must
receive sizable contributions from large left-right mixing in the stop
sector, possible for large $\tan\beta$ and/or large $|A_0|$. However,
in the CMSSM 
a large left-right mixing in the stop sector leads to a substantial
left-right mixing in the stau sector,  which in turn leads to a suppression of
the mass of the lighter stau.
For $m_0\sim 2-3$~TeV this results in the constraints $\tan\beta < 20$
and $A_0<-5$~TeV.  Finally, for $m_0$ of a few TeV, the staus are so
heavy that varying $\tan\beta$ or $|A_0|$ is not dangerous for the
DM relic density, so the Higgs boson mass measurement tends to push
$\tan\beta$ to higher values.

In Figure~\ref{CMSSMsigsip} the spin-independent direct detection cross
section \sigsip\ is shown as a function of the neutralino mass for
both the high $T_R$ scenario and for $T_R = 10$ GeV. As it is already known
\cite{Fowlie:2012}, a significant part of the $2\sigma$ credible
region in the high $T_R$ scenario can be tested in future one tonne
experiments. On the other hand, in the $T_R = 10$ GeV case prospects
for DM discovery are much worse. Only a small fraction of the allowed
region can be covered by Xenon1T; it is characterised by high $m_0$,
low $m_{1/2}$ and low $|A_0|$, where, according to
\cite{Kowalska:2014},  $\mu$ can be suppressed by the negative
$m_{H_u}^2(\textrm{SUSY})$ tending closer to zero, 
%%%%%%%%%%%%%%
\begin{equation}
-\mu^2 \simeq m_{H_u}^2(\textrm{SUSY}) \simeq 0.074\,m_0^2 -1.008\,m_{1/2}^2 - 0.080\,A_0^2 + 0.406\,m_{1/2}\,A_0 .
\label{CMSSMmHu2}
\end{equation}
%%%%%%%%%%%%%%
In this case the higgsino fraction of the bino-dominated DM goes up to even $5\%$.

\subsection{The NMSSM}
\label{sec:nmssm}

The superpotential of the MSSM contains a mass term $\mu \hat{H}_u
\hat{H}_d$ with a mass parameter $\mu$ of the order of the soft SUSY
breaking parameters. One therefore needs an explanation why $\mu$
should be much smaller than the other scales in the unbroken SUSY theory, such
as the unification scale or the Planck scale.

A simple and elegant solution to this `$\mu$-problem' consists in
replacing a mass term with a Yukawa-like interaction between the
chiral superfields $\hat{H}_u$, $\hat{H}_d$ and a chiral superfield
$\hat{S}$ which is a singlet of the SM gauge. Its scalar component can
acquire a VEV, thereby generating an effective $\mu$ term (see
\cite{Ellwanger:2009dp} for a review). This framework is called the
Next-to-Minimal Supersymmetric Standard Model (NMSSM).
The fermionic component of the singlet multiplet, the singlino,
carries no $SU(3)$ or electric charges, so it can mix with the other
four neutralinos. It is therefore possible that in the NMSSM a state
which is mostly singlino-like is the lightest of the neutral, non-SM,
$R$-parity-protected fermions and therefore a DM candidate.

The parameter space of the NMSSM contains three parameters absent in
the MSSM. They come from new terms in the superpotential $\lambda
\hat{S}\hat{H}_u\hat{H}_d+ 1/3!\,\kappa \hat{S}^3$ and from soft
SUSY breaking potential $1/3!\, A_\kappa S^3$ (the coefficient
$A_\lambda$ in the term $A_\lambda H_u H_d S$ of the soft SUSY
breaking potential is then determined in terms of other parameters,
including $\mu$ and $m_A$).  We therefore extend the numerical
analysis described in Section~\ref{secp10mssm} to accommodate these
three additional parameters; their ranges are given in Table~\ref{tabnmssm}. The
spectrum and the decay widths are calculated with {NMSSMTools 4.2.0}
\cite{Ellwanger:2005dv} and the high-$T_R$ relic density is obtained
from an appropriately extended {micrOMEGAs} code
\cite{Belanger:2005kh}.  Requiring perturbativity to hold  up to the
GUT scale requires $\lambda, \kappa \lesssim 0.7$; this justifies our
choice for the upper limit in the scan, but in practice there are
always additional constraints. The condition that the singlino is lighter
than the higgsino implies that $\kappa \lesssim \lambda/2 < 0.35$
and the requirement that the DM is made up of an almost pure singlino (a
parameter region which we focus on here) introduces an effective
upper limit $\lambda\lesssim 0.1$ for majority of points in the scan
-- this suppresses the respective off-diagonal entries in the neutralino
mass matrix. As a result one typically finds $\kappa < 0.05$.  The
effective upper limit on $A_{\kappa}$ comes from positivity of the
pseudoscalar mass matrix and it reads $A_{\kappa} \lesssim 0$.

%%%%%%%%%%%%%%
\begin{table}[t]
\centering
\begin{tabular}{|c|c|}
\hline 
Parameter & Range \\ 
\hline
\hline 
$SH_uH_d$ coupling & $0.001 < \lambda < 0.7$ \\ 
scalar cubic coupling & $0.001 < \kappa < 0.7$ \\ 
soft scalar $A$-term & $-12\,\rm{TeV} < A_\kappa < 12\,\rm{TeV}$ \\
\hline 
\end{tabular}
\caption{Additional parameters in the p13NMSSM and their ranges given
  at the SUSY scale. The nuisance parameters are the same as for the p10MSSM.} 
\label{tabnmssm} 
\end{table}
%%%%%%%%%%%%%%

%%%%%%%%%%%%%%
\begin{figure}[t]
\begin{center}
\includegraphics*[width=8cm,height=7cm]{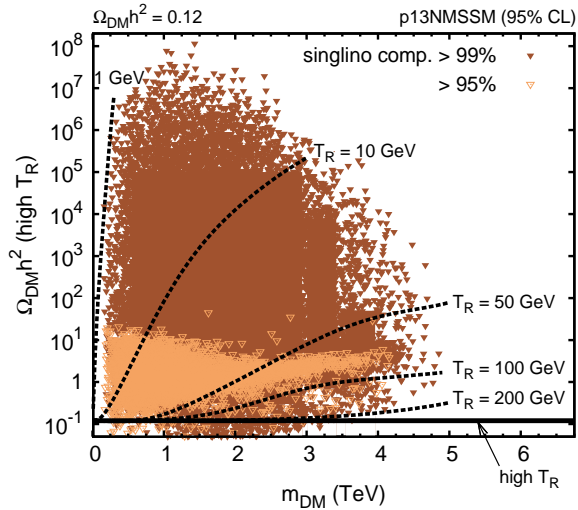}
\end{center}
\caption{Contours (black dotted) of constant $\Omega_{\textrm{DM}}h^2
 = 0.12$ for different values of the reheating
  temperature $T_R$ in the NMSSM with a singlino-like DM in a
  $\big(m_{\rm{DM}}, \Omega_{\rm{DM}}h^2(\textrm{high
  }T_R)\big)$. Solid black horizontal line corresponds to high $T_R$
  limit. Shown scan points correspond to credibility levels of 95\%;
  dark (light) brown triangles correspond singlino fraction $>99\%$
  (between 95\% and 99\%).
}
\label{f_nmssm}
\end{figure}
%%%%%%%%%%%%%%

The results of the scan projected onto the $(m_{\rm{DM}},
\Omega_{\rm{DM}}h^2(\textrm{high }T_R) )$ plane are shown in
Figure~\ref{f_nmssm}. The range of the high-$T_R$ singlino relic
density spans a few orders of magnitude, from $10^{-2}$ to $10^7$. The
largest values are $\sim4$ orders of magnitude larger than the largest
values that we obtained for the bino LSP.  This can be explained by
the fact that a nearly pure singlino interacts very weakly; it
annihilates mainly into scalar-pseudoscalar pairs (mainly
$H_2A_1$) with the associated couplings proportional to $\kappa$ or
$\lambda$. This dominant annihilation channel is characteristic of
scan points with $\Omega_{\rm{DM}}h^2(\textrm{high }T_R) > 10^5$.
Smaller values of $\Omega_{\rm{DM}}h^2(\textrm{high }T_R)$ require at
least a partial mass degeneracy between the singlino and a heavier
particle thus allowing coannihilations: these are mainly coannihilations
with the bino for $10^3<\Omega_{\rm{DM}}h^2(\textrm{high }T_R)<10^5$ and
coannihilations with the higgsino for $\Omega_{\rm{DM}}h^2(\textrm{high
}T_R)\sim10^2$.  We also find points with smaller values of
$\Omega_{\rm{DM}}h^2(\textrm{high }T_R)$, but they necessarily involve
special mass patterns which permit coannihilations with either 
higgsino, wino, stau/sneutrino, stop or gluino.

The lines of constant $\Omega_{\textrm{DM}}h^2 = 0.12$ for different
values of the reheating temperature $T_R$ shown in
Figure~\ref{f_nmssm} are the same as in Figure~\ref{MSSMclueTR} and we
arrive at a conclusion analogous to that of Section~\ref{secp10mssm},
namely that there exist vast regions of the parameter space of the
NMSSM with an almost pure singlino DM which have been so far
disregarded solely because of predicting too large a relic density;
however, with sufficiently low $T_R$ the relic density can be
suppressed enough to agree with the measured value and these regions
become phenomenologically viable.

%%%%%%%%%%%%%%

\section{Direct and/or cascade decays of the inflaton field}
\label{sec:decays}

We have so far made an implicit assumption that the inflaton field
$\phi$ is very heavy and, therefore, that the direct and cascade
decays of $\phi$ to DM species are negligible.  It is,
however, important to study the validity of this assumption for a
range of inflaton mass, as inflaton decays can give an additional, {\em non-thermal}
contribution to $\Omega_\chi h^2$. % after LOSP freeze-out. 
Our analysis follows here the model-independent approach used in
\cite{Gelmini:2006Feb,Gelmini:2006May}.

Direct and cascade decays of the inflaton field to superpartners
of SM particles correspond to an additional term in the Boltzmann equation
(\ref{Boltzmann}) for $n$, which is now given by,\footnote{The most
  important contribution from direct and cascade decays is associated
  with the period between the freeze-out of DM particles and the end
  of the reheating period when 
  $n$ becomes essentially equal to $n_\chi$.
}  
%%%%%
\beq
\frac{dn}{dt} = -3Hn - \langle\sigma v\rangle\big[n^2 -
(n^{eq})^2\big]+\frac{b}{m_\phi}\Gamma_{\phi}\rho_{\phi} \, ,
\label{DMBoltzminfl}
\eeq 
%%%%%
where $b$ describes the average number of DM particles produced per
inflaton decay described by the decay constant $\Gamma_\phi$
and $\rho_\phi$ denotes the inflaton energy density. 

%%%%%%%%%%%%%%
\begin{figure}[!t]
\begin{center}
\includegraphics*[width=8cm,height=7cm]{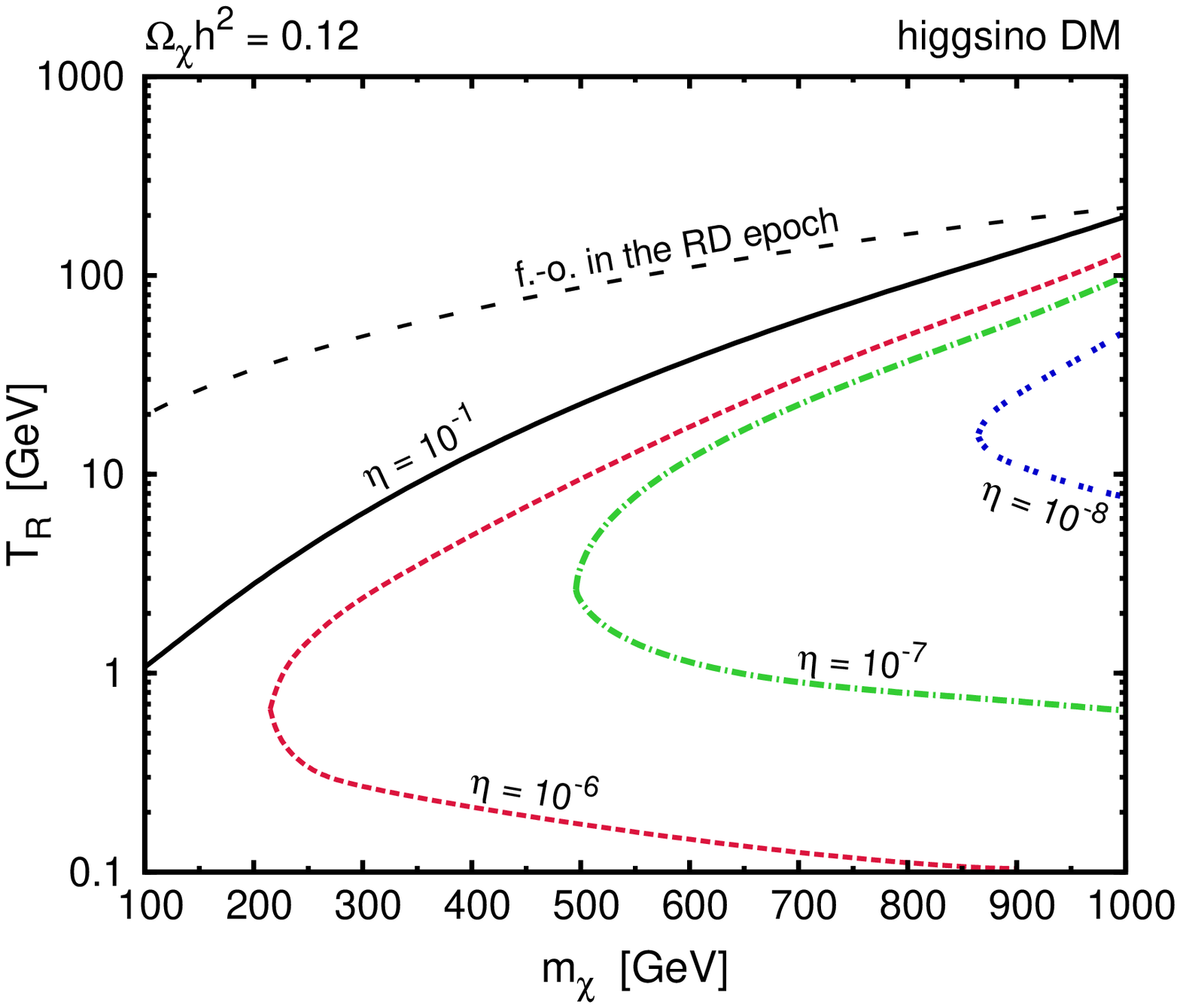}
\hfill
\includegraphics*[width=8cm,height=7cm]{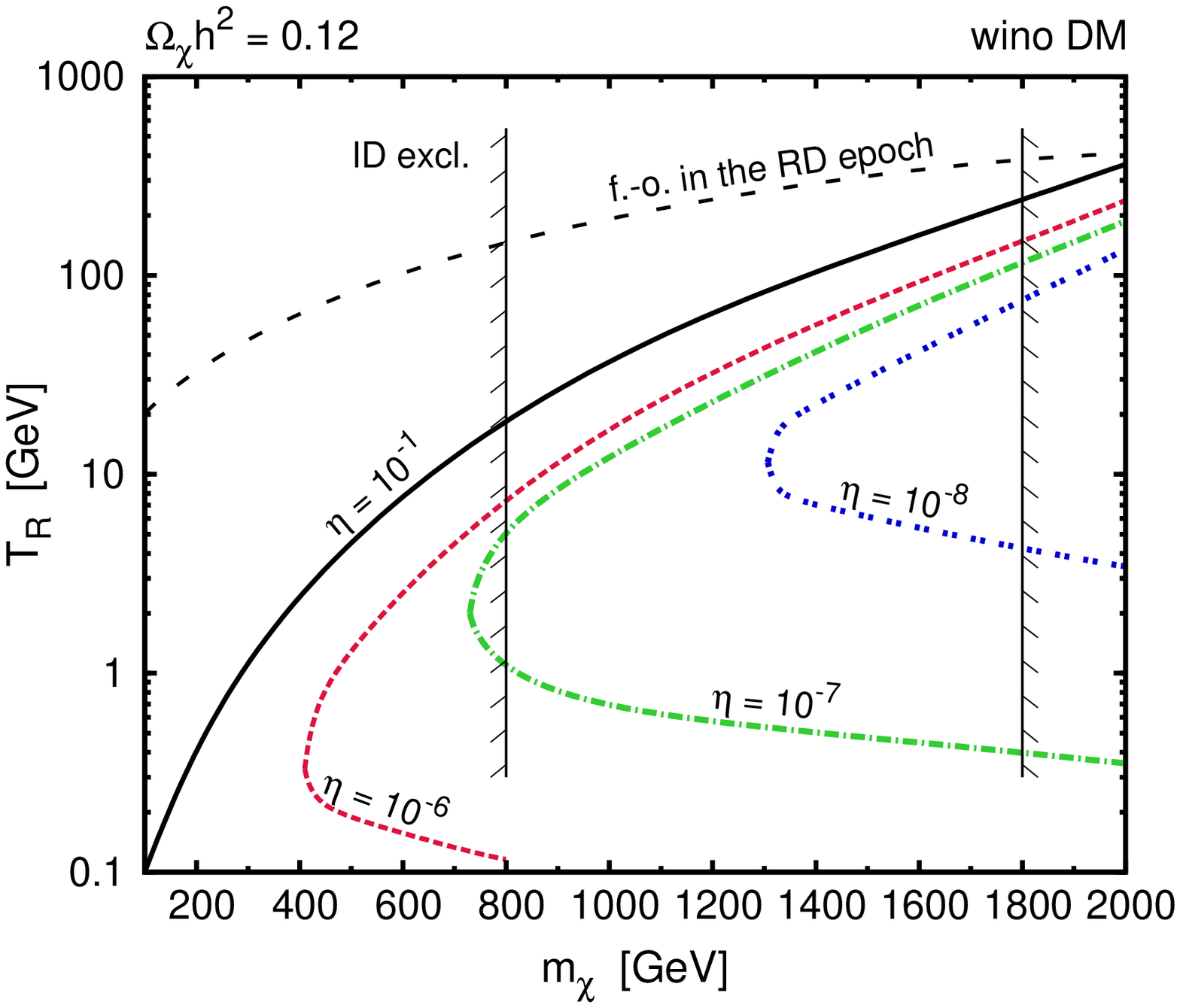}
\end{center}
\caption{Contours of constant $\Omega_\chi h^2 = 0.12$ in the
  ($m_\chi,T_R$) plane for different values of the dimensionless
  quantity $\eta = b\Big(100\,\textrm{TeV}/m_{\phi}\Big)$ for
higgsino (left panel) and wino (right panel) DM.  Solid black (dashed
red, dot-dashed green, dotted blue) lines correspond respectively to $\eta = 10^{-1}$
($10^{-6}$, $10^{-7}$, $10^{-8}$).  In the wino DM case we take indirect
detection limits following \cite{Hryczuk:2014}. For the reheating temperatures above thin dashed black lines the freeze-out of the DM particles occurs after the reheating period (i.e. in the RD epoch). The limit at $\sim 800$
GeV comes from antiprotons and the one around $1.8$ TeV from the
absence of a $\gamma$-ray line feature towards the Galactic Center.}
\label{infldec}
\end{figure}
%%%%%%%%%%%%%%

We present our results in Figure~\ref{infldec} in the ($m_\chi,T_R$)
plane in terms of the dimensionless quantity 
\mbox{$\eta = b\cdot(100\,\textrm{TeV}/m_{\phi})$} 
for higgsino (left panel) and wino (right panel) DM.

The relic density of DM in this case is a sum of the thermal and the
non-thermal components. The thermal production with a low reheating
temperature has been studied in Section~\ref{sec:fo} and shown to
be an increasing function of $T_R$. 
On the other hand, the magnitude of the non-thermal component may depend, for fixed $\eta$ and $m_\chi$, on the reheating temperature in a non-monotonic way, as discussed in detail in \cite{Gelmini:2006May}. When $T_R$ is sufficiently low, non-thermal production leads to $\Omega_\chi \sim T_R$, while for larger reheating temperature DM relic density goes down with increasing $T_R$.
 As a consequence, each curve
corresponding to fixed relic density $\Omega_\chi h^2 = 0.12$ and
fixed $\eta$ in Figure~\ref{infldec} is C-shaped.
For the upper branch of each curve, corresponding to larger values of $T_R$, the correct relic
density is obtained for such values of $m_\chi$ that freeze-out occurs
only slightly earlier than the end of the reheating
period.\footnote{Note that this happens at temperatures somewhat lower
  than $T_R$, as the reheating temperature {\em does not} mark the end of the reheating period.}
As $m_\chi$ increases required values of the $T_R$ become larger and finally reach the level at which freeze-out occurs after the reheating period, i.e., in the RD epoch, and therefore direct and cascade decays of the inflaton field play no role in determining $\Omega_\chi$.

The additional, non-thermal contribution to the DM relic abundance can help reconcile with the measured value these regions of the
MSSM parameter space for which $\Omega_\chi h^2$ is otherwise too low
even at high $T_R$. Examples of such cases include the higgsino with
mass below 1~TeV or wino with mass below 2~TeV, shown in
Figure~\ref{infldec}. For sufficiently large values of $\eta$, one can even
generate too much DM from inflaton decays; this upper bound on $\eta$
can be translated into a lower bound on the inflaton mass for which
the direct production is negligible even for a branching ratio
$\mathrm{BR}(\phi\to\textrm{superpartners})\sim\mathcal{O}(1)$. In
particular, for $\eta <10^{-9}$ we obtain no significant
non-thermal production of DM particles.  This value corresponds to the
inflaton mass of $m_\phi =b\cdot 10^{13}\,\mathrm{GeV}$, which for
typical values of $b\sim\mathcal{O}(10^3)$ \cite{Kurata:2012nf},
points towards inflaton mass close to the unification scale.\footnote{The inflaton mass during
  reheating, when the inflaton field oscillates coherently around the
  minimum of the potential, can significantly differ from the inflaton
  mass parameter during inflation, usually quantified by slow-roll
  parameters. We use the former. } 

%%%%%%%%%%%%%%

\section{Gravitino dark matter with low reheating temperature}
\label{sec:grav}

Many of the considerations presented in Section~\ref{sec:neudm} can be
applied to another theoretically motivated scenario where the DM is
made up of the gravitino $\widetilde{G}$, the supersymmetric partner
of the graviton, assuming that $\widetilde{G}$ is lighter than all the
superpartners of the SM particles.

Unlike the neutralino, for a sufficiently large mass the gravitino is
not a thermal relic. Its abundance $\Omega_{\widetilde G}h^2$ receives
contributions from at least two sources: the thermal component
$\Omega_{\widetilde G}^\mathrm{TP}h^2$ is produced in scatterings and
decays in the thermal plasma \cite{Bolz:2000,Pradler:2006,
  Rychkov:2007}, while the nonthermal component $\Omega_{\widetilde
  G}^\mathrm{NTP}h^2$ results from late decays of quasi-stable relic
LOSPs after they freeze out \cite{Ellis:1984,Ellis:1992}. Since
$\Omega_{\widetilde G}^\mathrm{TP}h^2$ is proportional to $T_R$, for
$T_R\ll 10^6$~GeV and $m_{\widetilde G}\simgt 1$~GeV this component is
much smaller than the measured value of the relic density, hence at 
low $T_R$ it is the nonthermal component of gravitino DM that
is dominant, and the gravitino abundance can be related to the LOSP
abundance by
%%%%%%%%%%%%%%
\begin{equation}
%lr \Omega_{\widetilde{G}}h^2(\textrm{low }T_R) \simeq
%lr \Omega_{\widetilde{G}}^{\textrm{NTP}}h^2(\textrm{low }T_R) =
\Omega_{\widetilde{G}}h^2 \simeq
\Omega_{\widetilde{G}}^{\textrm{NTP}}h^2 =
\frac{m_{\widetilde{G}}}{m_{\textrm{LOSP}}}\,\Omega_{\textrm{LOSP}}h^2 .
\label{gravreldens}
\end{equation}
%%%%%%%%%%%%%%
Long after they have frozen out, during or after BBN, the LOSPs decay
into gravitinos and SM particles, thus initiating hadronic and
electromagnetic cascades which can affect light element abundances (see
e.g. \cite{Jedamzik:2006,Iocco:2008}) and potentially lead to a
violation of current observational limits.

Here we analyze the viability of the gravitino DM with low reheating
temperatures, making use of the results of the scan described in Section
\ref{secp10mssm} with an additional assumption that the gravitino is
lighter than any of the superpartners of the SM particles and without
requiring that the LOSP is neutral which allows the LOSP to be a
neutralino (bino, wino or higgsino) or a slepton (a charged slepton or,
with large enough splitting between right and left soft stau masses
\cite{Roszkowski:2012}, a sneutrino).  We follow
Ref.~\cite{Jedamzik:2006} for the implementation of BBN constraints, which
mainly depend on the LOSP mass $m_\mathrm{LOSP}$ and abundance
$\Omega_{\textrm{LOSP}}h^2$, as well as on the LOSP hadronic branching
ratio $B_{\rm{h}}$.  We calculate $\Omega_{\textrm{LOSP}}h^2$ as
described in Section \ref{secp10mssm} and for $B_{\rm{h}}$ we use
existing results for neutralinos \cite{Covi:2009}, sneutrinos
\cite{Kanzaki:2006} and charged sleptons \cite{Steffen:2006}.

\begin{figure}[!t]
\begin{center}
\includegraphics*[width=8cm,height=7cm]{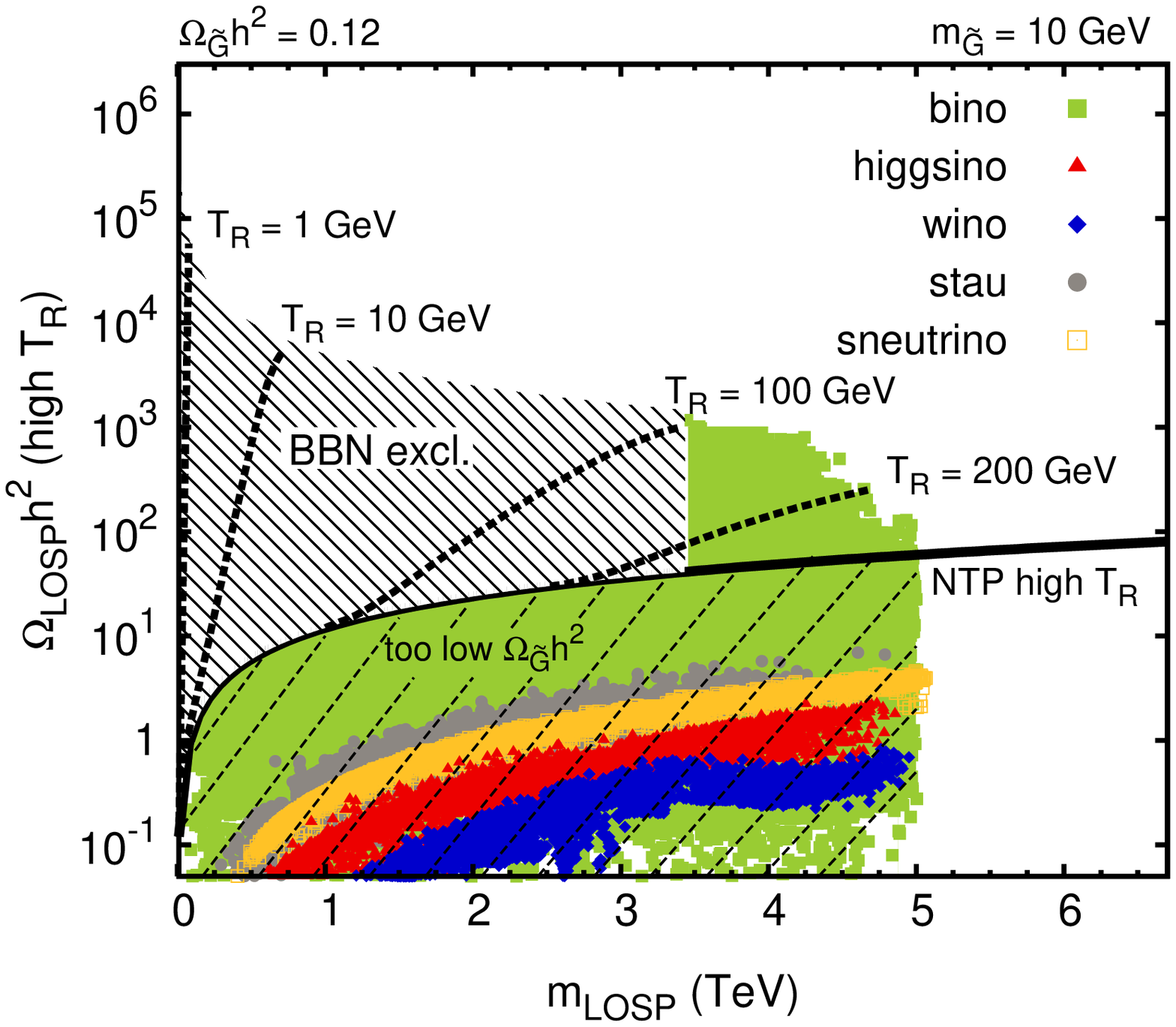}
\hfill
\includegraphics*[width=8cm,height=7cm]{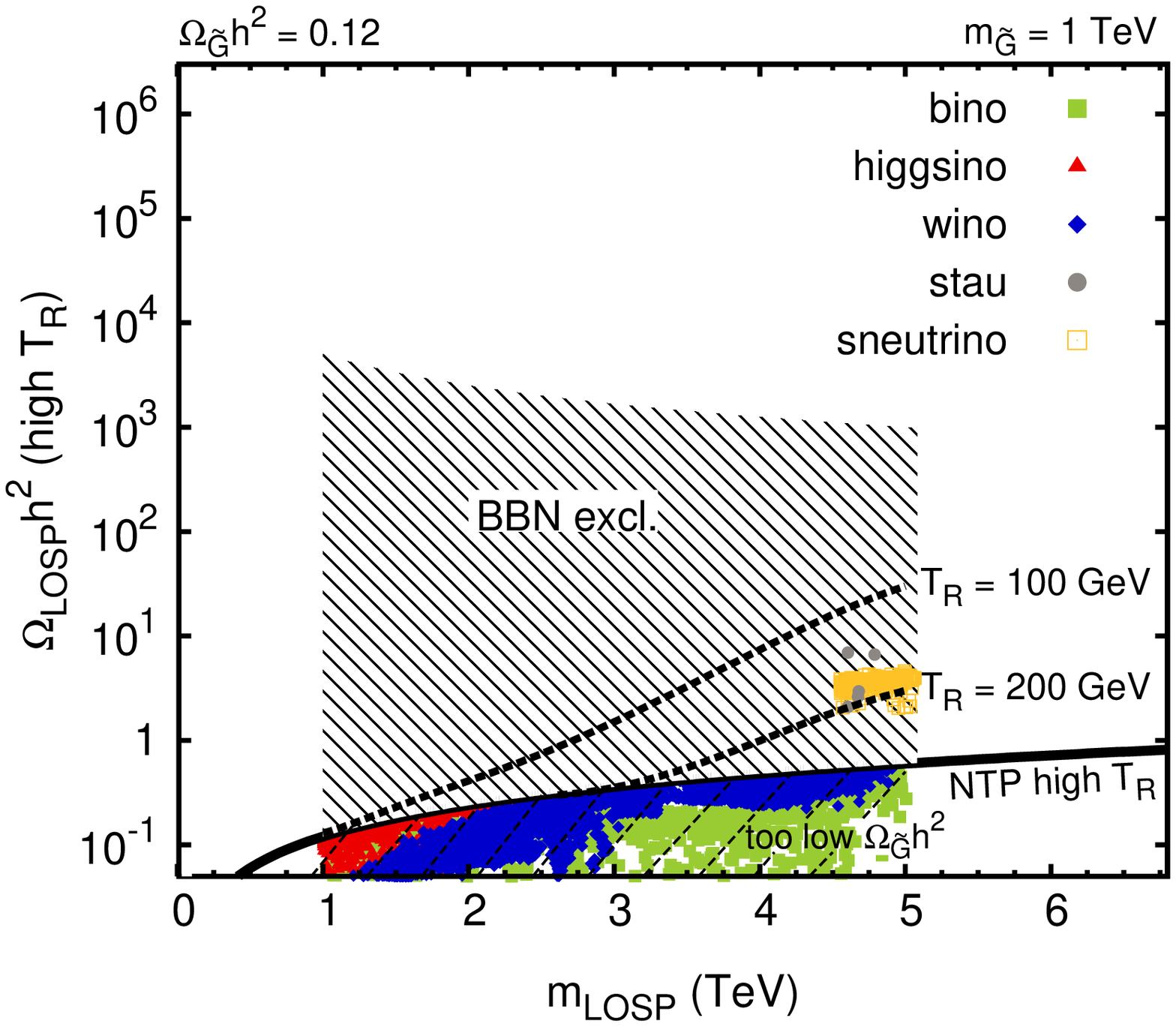}
\end{center}
\caption{Contours of constant $\Omega_{\widetilde{G}}h^2 = 0.12$ for different values of the reheating temperature
  $T_R$ and for $m_{\widetilde{G}} = 10$ GeV and 1~TeV in the p10MSSM
  with BBN constraints imposed. Color coding as in
  Figure~\ref{MSSMclueTR}. }
\label{MSSMclueTRgrav}
\end{figure}

Typical results for $m_{\widetilde{G}} = 10$~GeV and 1~TeV obtained in
the p10MSSM are given in Figure~\ref{MSSMclueTRgrav}.  We fix the
gravitino abundance at the observed value, relate it to the LOSP relic
density through (\ref{gravreldens}) and then find the corresponding
reheating temperature with the procedure described in Section
\ref{sec:neudm}.  Similarly as in Section \ref{secp10mssm}, we present
the results in the $(m_\mathrm{LOSP},
\Omega_{\mathrm{LOSP}}h^2(\textrm{high }T_R))$ plane.  As one could
expect from eq.~(\ref{gravreldens}), the line corresponding to the
correct gravitino DM abundance in high-$T_R$ case is not horizontal,
as it was the case for neutralino DM.  Below the line the gravitino
abundance is lower than the observed value and, in the absence of
thermally produced component,\footnote{In the cases discussed here the
  value of $T_R$ below which $\Omega_{\widetilde G}^\mathrm{TP}h^2$ is
  negligible is a few orders of magnitude larger than a `low' value of
  $T_R$ and it makes sense to consider a high-$T_R$ limit without the
  thermally generated component of gravitino DM.}  such points are not
viable.  We note that for the sneutrino LOSP it is mass degenerate with
the lighter (left) stau, thus coannihilations do play an important
role here.  This typically makes
$\Omega_{\mathrm{LOSP}}h^2(\textrm{high }T_R)$ smaller for the sneutrino
LOSP than for the (usually right) stau LOSP.

%%%%%%%%%%%%%%
%%%%%%%%%%%%%%

In the low $T_R$ regime, as long as $m_{\widetilde{G}} \lesssim 100$~GeV, the bino as the LOSP is the only possibility for gravitino DM.
In this case, however, $\Omega_{\mathrm{LOSP}}h^2(\textrm{high }T_R)$
typically exceeds unity and $B_{\rm h}\sim 1$; hence, in order to
avoid bounds from the BBN one can simply require the LOSP lifetime to
be $\simlt 0.1$~s, which leads to \cite{Covi:2009}
%%%%%%%%%%%%%%
\begin{equation}
m_\mathrm{LOSP} \simgt 1400\, \left( \frac{m_{\widetilde G}}{\mathrm{GeV}}\right)^{2/5}\,\mathrm{GeV}\,,
\end{equation}
%%%%%%%%%%%%%%
which is consistent with the results shown in the left panel of
Figure~\ref{MSSMclueTRgrav}. The interpretation of this bound is very
simple: the LOSP  number density is so large that the particle must
decay before BBN in order not to affect its successful predictions;
because of 
the lifetime-mass dependence, this places a stringent lower bound on the
LOSP mass. While at low $T_R$ one can suppress the LOSP number density and
alleviate BBN constraints, with a small $m_{\widetilde
  G}/m_\mathrm{LOSP}$ in (\ref{gravreldens}) this would lead to too
small gravitino abundance. 

\begin{figure}[!t]
\begin{center}
\includegraphics*[width=7.5cm,height=7cm]{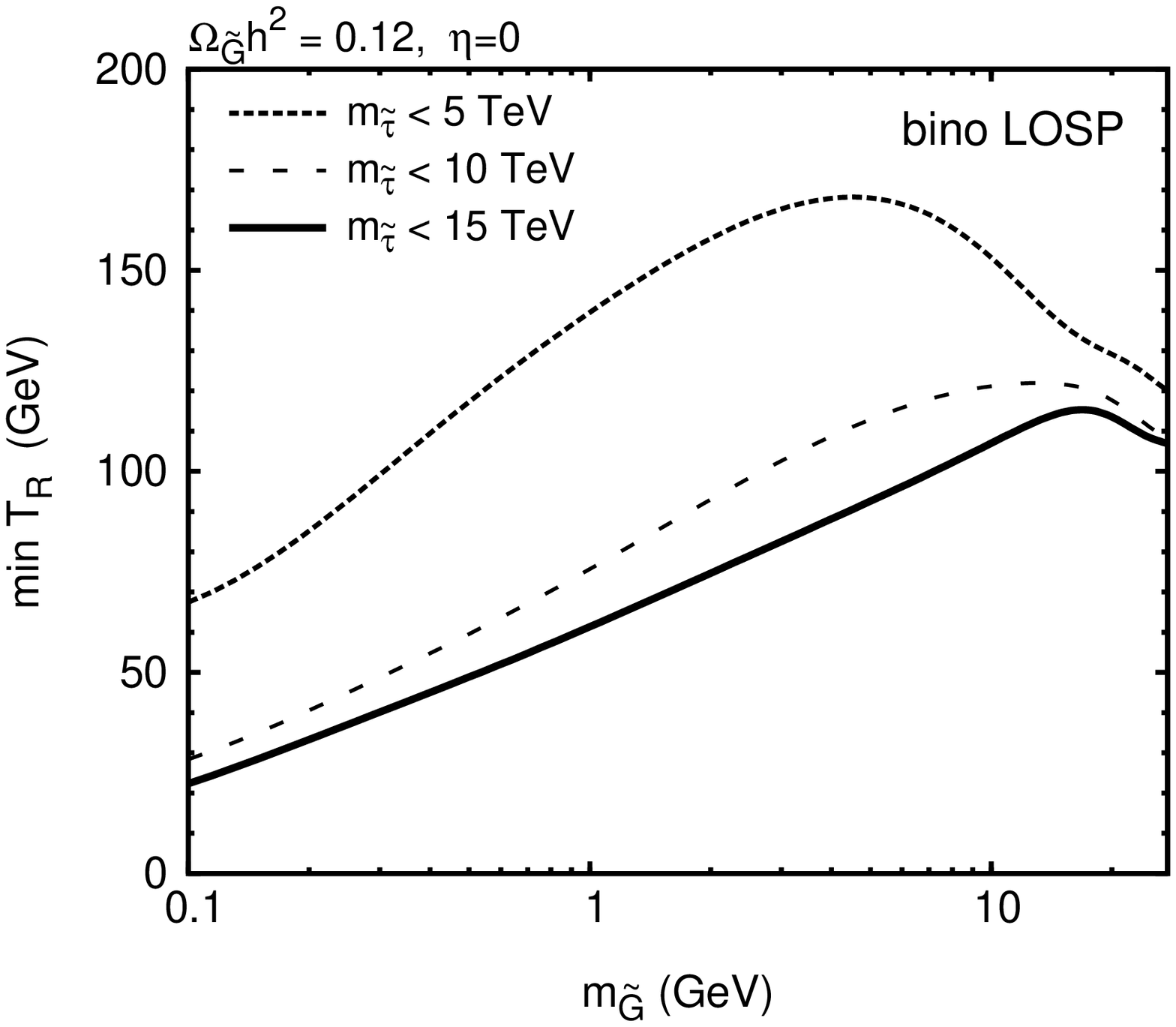}
\hfill
\includegraphics*[width=7.5cm,height=7cm]{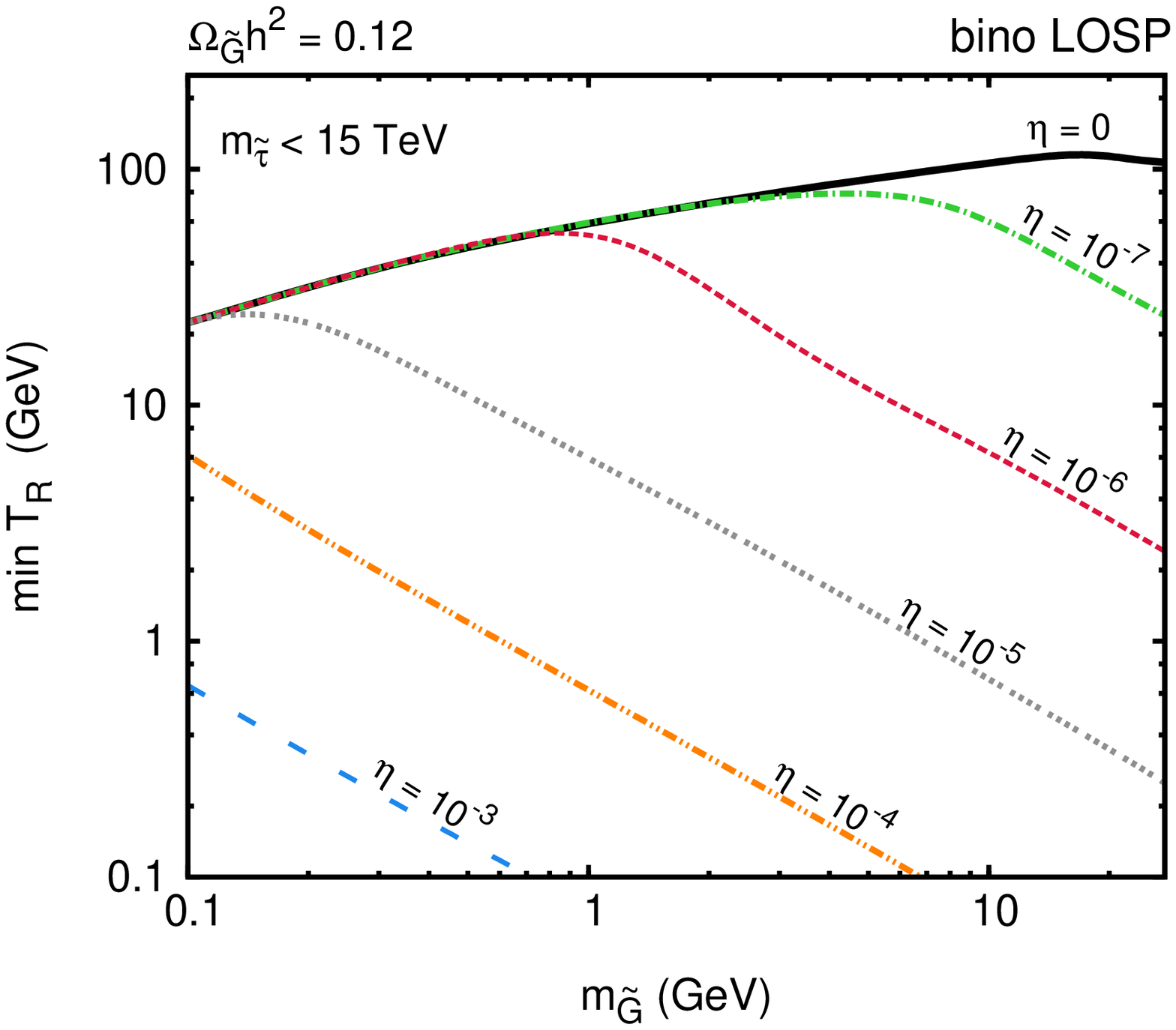}
\end{center}
\caption{Lower bounds on $T_R$ as a function of $m_{\widetilde G}$ for
  gravitino DM with a bino LOSP.
  On the left panel, direct and cascade decays of the inflaton field to bino
  are neglected and three choices of the maximal
  stau mass $m_{\widetilde \tau}=5,\,10$ and~15~TeV are shown.
  On the right panel, the effects of the inclusion of the direct and cascade decays of the inflaton
  is shown for different values of $\eta$ and fixed $m_{\widetilde \tau}=15$~TeV.
   }
\label{ftrb}
\end{figure}

On the other hand, it follows from
Figure~\ref{MSSMclueTRgrav} that a lower bound on $m_\mathrm{LOSP}$ can
be translated into a lower bound on $T_R$. We show such bounds in
Figure~\ref{ftrb} 
%(left panel) 
as a function of the gravitino mass with and without efficient direct and cascade decays of the inflaton field to bino.  
As we argued in Section \ref{secp10mssm},  the upper boundary of the
points in Figure~\ref{MSSMclueTRgrav} corresponds to the maximum
value of the stau mass, so the lower limits on $T_R$ with bino LOSP are
presented for three maximum values of the stau mass: 5,~10~and~15~TeV.

%%%%%%%%%%%%%%
%%%%%%%%%%%%%%

A qualitatively different picture emerges when $m_{\widetilde
  G}\gtrsim 100$~GeV. The LOSP lifetime is then so large that the BBN
bounds can only be evaded when $B_{\rm h}$ is small and
$m_\mathrm{LOSP} \gtrsim 1$ TeV with the number density reduced
because of low $T_R$.  This is, however, only possible for the
sneutrino and, very rarely, for the stau LOSP \cite{Roszkowski:2004jd,Cerdeno:2005eu}, as
presented in the right panel of Figure~\ref{MSSMclueTRgrav} for
$m_{\widetilde G}=1$~TeV.\footnote{In our case the
  stau LOSP scenario is only slightly constrained by the possibility
  of forming bound states with nuclei
  \cite{Pospelov:2006sc,Jedamzik:2007qk,Kawasaki:2007xb,Kawasaki:2008qe}
  due to a relatively low stau lifetime; for the same reason CMB
  constraint \cite{Hu:1992dc,Hu:1993gc,Roszkowski:2004jd} plays no
  role here, either. A recent analysis of a scenario with gravitino DM
  and stau LOSP in the context of the LHC searches can be found in
  \cite{Heisig:2013sva}.}  Hence also for $m_{\widetilde G}\gtrsim
100$~GeV we find a lower bound 
$T_R\gtrsim 150$~GeV. This is true if direct and cascade decays of the inflaton field to the LOSP 
can be neglected; otherwise, 
the lower limit on $T_R$ becomes less severe, similarly to the bino LOSP case.
 
If one assumes gaugino mass unification at the GUT scale, then the
lower limit on the chargino mass from collider searches,
$m_{\chi_1^{\pm}} > 94$ GeV \cite{PDG:2014}, can be translated into a
lower limit on the lightest neutralino mass $m_{\chi} > 46$ GeV. This
in turns implies in our p10MSSM scan $m_{\widetilde{G}} \gtrsim 0.1$
MeV, where we assume soft scalar masses not to be greater than $\sim
15$ TeV and $T_R$ low enough so that the gravitino is produced only in NTP. For much lighter gravitinos, in the keV mass range, the correct abundance can be
obtained by thermal production for reheating temperature even of the
order of a few hundred GeV (see, e.g., \cite{Bomark:2014}).

It is important to note that the additional contribution to the LOSP relic density resulting from direct and/or cascade decays of the inflaton
 allows one to consider lower values of the reheating temperature in gravitino DM scenario. In such a case, the lower limit on $T_R$ becomes less severe, as it is illustrated in the right panel of Figure~\ref{ftrb} for the bino LOSP; the same is true for the slepton LOSP.

\section{Conclusions}
\label{sec:conc}

Motivated by the observation that in scenarios with a low reheating
temperature  DM relic density is reduced with respect
to the standard high-$T_R$ case, in this paper we studied the impact of
assuming low $T_R$ on the phenomenologically
favored regions of the (C)MSSM and the NMSSM with the singlino DM.  We
considered two distinct DM candidates: the LOSP and the gravitino.

In the case of the LOSP we found that, at low $T_R$ large regions of
the parameter space open up which are normally considered
excluded because of too large a relic density.  With $T_R$ in the
range $100-200$~GeV, the DM can be the bino (coannihilating with
staus), the heavy ($\simgt 3.5$~TeV) wino or the higgsino if it is not
lighter than about 1~TeV. For $T_R=\mathcal{O}(10)$~GeV the allowed
regions of the parameter space mainly correspond to a bino-like
neutralino in the bulk region, with a small fraction of solutions
having a higgsino admixture of a few per cent. Similarly, in the
singlino-dominated region of the NMSSM, when $T_R$ is less than about
$200$~GeV, large regions open up where at high $T_R$ the relic density
can be very high.

If DM consists of nonthermally produced
gravitinos only, then the relic abundance of LOSPs decaying into gravitinos must
be greater than the observed dark matter abundance. Since the effect
of low $T_R$ is to reduce the LOSP relic abundance, $T_R$ cannot be
too low. In this case we obtain lower bounds on $T_R$ by combining
the assumed generous upper bounds on the superpartner masses of a few TeV (which reflects
our view that SUSY should not lead to a too severe hierarchy problem)
and the BBN constraints.  For bino (slepton) LOSP, we find the bound
on $T_R$ of the order of 100~GeV for the gravitino mass in the range
$0.1-10$ ($10^2-10^3$)~GeV. These limits are alleviated when significant direct and cascade decays of the inflaton field to the LOSP are present.

\section*{Acknowledgements}
We are grateful to K.~Y.~Choi for comments.  This work has been funded
in part by the Welcome Programme of the Foundation for Polish Science.
The work of KT was supported in part by the Polish National Science
Centre grant N N202 167440 and in part by grant IP2011 056971 from the
Ministry of Science and Higher Education. LR is also supported in part
by a STFC consortium grant of Lancaster, Manchester, and Sheffield
Universities. The use of the CIS computer cluster at
the National Centre for Nuclear Research is gratefully acknowledged.

\end{document}

%% file: macros.tex
\newcommand{\newc}{\newcommand*}

\long\def\begincomment#1\endcomment{%
        \begingroup\sf\baselineskip12pt#1\endgroup}

\newc{\etal}{\textrm{et al.}} 
\newc{\eg}{\textrm{e.g.}} 
\newc{\ie}{\textrm{i.e.}}
\newc{\etc}{\textrm{etc}}
\newc\vs{\textrm{vs.}}
\newc{\cl}{\rm {C.L.}}
\newc{\ev}{\ensuremath{\,\mathrm{eV}}}
\newc{\kev}{\ensuremath{\,\mathrm{keV}}}
\newc{\mev}{\ensuremath{\,\mathrm{MeV}}}
\newc{\gev}{\ensuremath{\,\mathrm{GeV}}}
\newc{\tev}{\ensuremath{\,\mathrm{TeV}}}
\newc{\MeV}{\mev} 
\newc{\TeV}{\tev}
\newc{\invpb}{\ensuremath{/\text{pb}}}
\newc{\invfb}{\ensuremath{/\text{fb}}}
\newc\nb{\ensuremath{\,\mathrm{nb}}} \newc\pb{\ensuremath{\,\mathrm{pb}}} \newc\fb{\ensuremath{\,\mathrm{fb}}}
\newc\pc{\ensuremath{\,\mathrm{pc}}}
\newc\kpc{\ensuremath{\,\mathrm{kpc}}}
\newc\mpc{\ensuremath{\,\mathrm{Mpc}}}
\newc\ps{\ensuremath{\,\mathrm{ps}}} 
\newc\cmeter{\ensuremath{\,\mathrm{cm}}} 
\newc\meter{\ensuremath{\,\mathrm{m}}} 
\newc\kmeter{\ensuremath{\,\mathrm{km}}}
\newc\second{\ensuremath{\,\mathrm{s}}}
\newc\msecond{\ensuremath{\,\mathrm{ms}}}
\newc\nsecond{\ensuremath{\,\mathrm{ns}}}
\newc\psecond{\ensuremath{\,\mathrm{ps}}}
\newc{\chisqmin}{\ensuremath{\chi^2_{\mathrm{min}}}}
\newc{\Delchisq}{\ensuremath{\Delta\chi^2}}
\newc{\chisq}{\ensuremath{\chi^2}}
\newc{\like}{\ensuremath{\mathcal{L}}}
\newc\lsim{\ensuremath{\mathrel{\rlap{\lower4pt\hbox{\hskip1pt$\sim$}}\raise1pt\hbox{$<$}}}}
\newc\gsim{\ensuremath{\mathrel{\rlap{\lower4pt\hbox{\hskip1pt$\sim$}}\raise1pt\hbox{$>$}}}}
\newc{\VEV}[1]{\ensuremath{\langle #1 \rangle}}
\newc{\dl}{\ensuremath{\stackrel{\leftarrow}{D}}}
\newc{\dr}{\ensuremath{\stackrel{\rightarrow}{D}}}
\newc{\bcenter}{\begin{center}}    \newc{\ecenter}{\end{center}}
\newc{\bfl}{\begin{flushleft}}    \newc{\efl}{\end{flushleft}}
\newc{\bfr}{\begin{flushright}}    \newc{\efr}{\end{flushright}}

\newc{\bi}{\begin{itemize}}
\newc{\ei}{\end{itemize}}
\newc{\bed}{\begin{description}}
\newc{\eed}{\end{description}}
\newc{\ben}{\begin{enumerate}}
\newc{\een}{\end{enumerate}}

\newc{\be}{\begin{equation}}
\newc{\ee}{\end{equation}}
\newc{\bea}{\begin{eqnarray}}
\newc{\eea}{\end{eqnarray}}
\newc{\ra}{\rightarrow}
\newc{\alphas}{\ensuremath{\alpha_s}}
\newc{\alphatwo}{\ensuremath{\alpha_2}}
\newc{\alphaone}{\ensuremath{\alpha_1}}
\newc{\alphai}[1]{\ensuremath{\alpha_{#1}}}
\newc{\alphaem}{\ensuremath{\alpha_{\mathrm{em}}}}
\newc{\alphaeff}{\ensuremath{\alpha_{\mathrm{eff}}}}
\newc{\sineff}{\ensuremath{\sin \theta_{\mathrm{eff}}}}
\newc{\sinsqeff}{\ensuremath{\sin^2 \theta_{\mathrm{eff}}}}
\newc{\dalphahad}{\ensuremath{\Delta \alpha_{\mathrm{had}}}}
\newc{\yt}{\ensuremath{h_t}} \newc{\yb}{\ensuremath{h_b}} \newc{\ytau}{\ensuremath{h_{\tau}}}
\newc\mz{\ensuremath{M_Z}} 
\newc\mw{\ensuremath{m_W}}
\newc\mZ{\mz}        \newc\mW{\mw}
\newc\mhsm{\ensuremath{ m_{H_{\mathrm{SM}}}}}
\newc{\mtop}{\ensuremath{ m_t}}               \newc{\mtpole}{\ensuremath{ M_t}}
\newc{\mbottom}{\ensuremath{ m_b}} 
\newc{\mtau}{\ensuremath{ m_{\tau}}}
\newc{\mt}{\mtpole}
\newc{\mb}{\mbottom} 
\newc{\rgg}{\ensuremath{R_{h}(\gamma\gamma)}}
\newc{\rzz}{\ensuremath{R_{h}(ZZ)}}
\newc{\rtwogg}{\ensuremath{R_{h_2}(\gamma\gamma)}}
\newc{\rtwozz}{\ensuremath{R_{h_2}(ZZ)}}
\newc{\ronegg}{\ensuremath{R_{h_1}(\gamma\gamma)}}
\newc{\ronezz}{\ensuremath{R_{h_1}(ZZ)}}
\newc{\rsiggg}{\ensuremath{R_{h_\textrm{sig}}(\gamma\gamma)}}
\newc{\rsigzz}{\ensuremath{R_{h_\textrm{sig}}(ZZ)}}
\newc{\llbar}{\ensuremath{\ell\bar{\ell}}}
\newc{\tauptaum}{\ensuremath{ \tau^+\tau^-}}
\newc{\qqbar}{\ensuremath{ q\bar{q}}} \newc{\ppbar}{\ensuremath{ p\bar{p}}}
\newc{\bbbar}{\ensuremath{ b\bar{b}}} \newc{\ttbar}{\ensuremath{ t\bar{t}}}
\newc{\ffbar}{\ensuremath{ f\bar{f}}} \newc{\tautaubar}{\ensuremath{ \tau\bar{\tau}}}
\newc{\mchi}{\ensuremath{m_{\chi}}}
\newc{\squark}{\ensuremath{\tilde{q}}}
\newc{\slepton}{\ensuremath{\tilde{l}}}
\newc{\gluino}{\ensuremath{\tilde{g}}} 
\newc{\mgluino}{\ensuremath{{m_{\gluino}}}}
\newc{\tone}{\ensuremath{{\tilde{t}_1}}}
\newc{\sthw}{\ensuremath{ \sin\theta_W}}              \newc{\cthw}{\ensuremath{\cos\theta_W}}
\newc{\tanthw}{\ensuremath{ \tan\theta_W}}              \newc{\cotthw}{\ensuremath{\cot\theta_W}}
\newc{\ssqthw}{\ensuremath{\sin^2 \theta_W}}
\newc{\msbar}{\ensuremath{\overline{MS}}} \newc{\drbar}{\ensuremath{\overline{DR}}}
\newc{\mtmtsmmsbar}{\ensuremath{ m_t(m_t)^{\msbar}_{{\mathrm{SM}}}}}
\newc{\mtmtsmdrbar}{\ensuremath{ m_t(m_t)^{\drbar}_{{\mathrm{SM}}}}}
\newc{\mtmtmssmdrbar}{\ensuremath{ m_t(m_t)^{\drbar}_{{\mathrm{SUSY}}}}}
\newc{\mbmbmsbar}{\ensuremath{ m_b(m_b)^{\msbar} }}
\newc{\mbmbsmmsbar}{\ensuremath{ m_b(m_b)^{\msbar}_{{\mathrm{SM}}}}}
\newc{\mbmzsmmsbar}{\ensuremath{ m_b(\mz)^{\msbar}_{{\mathrm{SM}}}}}
\newc{\mbmzsmdrbar}{\ensuremath{ m_b(\mz)^{\drbar}_{{\mathrm{SM}}}}}
\newc{\mbmzmssmdrbar}{\ensuremath{ m_b(\mz)^{\drbar}_{{\mathrm{SUSY}}}}}
\newc{\mtaumzsmmsbar}{\ensuremath{ m_{\tau}(\mz)^{\msbar}_{{\mathrm{SM}}}}}
\newc{\mtaumzsmdrbar}{\ensuremath{ m_{\tau}(\mz)^{\drbar}_{{\mathrm{SM}}}}}
\newc{\mtaumzmssmdrbar}{\ensuremath{ m_{\tau}(\mz)^{\drbar}_{{\mathrm{SUSY}}}}}
\newc{\alphasmzms}{\ensuremath{\alpha_s(M_Z)^{\overline{MS}}}}
\newc{\alphaimzms}[1]{\ensuremath{\alpha_{#1}(M_Z)^{\overline{MS}}}}
\newc{\alphaemmz}{\ensuremath{\alpha_{\mathrm{em}}(M_Z)^{\overline{MS}}}}
\newc{\mzero}{\ensuremath{{m_0}}}
\newc{\mhalf}{\ensuremath{ m_{1/2}}}
\newc{\tanb}{\ensuremath{\tan\beta}}
\newc{\azero}{\ensuremath{ A_0}}
\newc{\bzero}{\ensuremath{ B_0}}
\newc{\signmu}{\ensuremath{\rm{sgn}\,\mu}}
\newc{\mueff}{\ensuremath{\mu_{\rm{eff}}}}
\newc{\lam}{\ensuremath{{\lambda}}}
\newc{\kap}{\ensuremath{{\kappa}}}
\newc{\alam}{\ensuremath{{A_{\lambda}}}}
\newc{\akap}{\ensuremath{{A_{\kappa}}}}
\newc{\hs}{\ensuremath{ H_s}}      
\newc{\mhs}{\ensuremath{ m_{H_s}}} 
\newc{\mgut}{\ensuremath{ M_{\rm GUT}}}
\newc{\mplanck}{\ensuremath{ M_{\rm P}}}      \newc{\mpl}{\ensuremath{ M_{\rm Pl}}}
\newc{\msusy}{\ensuremath{ M_{\rm SUSY}}}      \newc{\ms}{\ensuremath{ M_{\rm S}}}
 \newc{\mhl}{\ensuremath{m_\hl}} 
 \newc{\mhone}{\ensuremath{m_{h_1}}} 
 \newc{\mhtwo}{\ensuremath{m_{h_2}}} 
 \newc{\mglu}{\ensuremath{m_{\tilde g}}} 
 \newc{\mul}{\ensuremath{m_{\tilde{u}_L}}} 
 \newc{\mtone}{\ensuremath{m_{\tilde{t}_1}}} 
 \newc{\ma}{\ensuremath{m_A}} 
 \newc{\maone}{\ensuremath{m_{a_1}}} 
 \newc{\matwo}{\ensuremath{m_{a_2}}}
 \newc{\hone}{\ensuremath{h_1}}
 \newc{\htwo}{\ensuremath{h_2}}
 \newc{\aone}{\ensuremath{a_1}}
 \newc{\atwo}{\ensuremath{a_2}}
 \newc{\mhu}{\ensuremath{ m_{H_u}}}       
 \newc{\mhd}{\ensuremath{ m_{H_d}}}
 \newc{\mhusq}{\ensuremath{ m_{H_u}^2}}       
 \newc{\mhdsq}{\ensuremath{ m_{H_d}^2}}
 \newc{\mhuew}{\ensuremath{ m^{\ast}_{H_u}}}       
 \newc{\mhdew}{\ensuremath{ m^{\ast}_{H_d}}}
 \newc{\mhuewsq}{\ensuremath{ m^{\ast\, 2}_{H_u}}}       
 \newc{\mhdewsq}{\ensuremath{ m^{\ast\, 2}_{H_d}}}
 \newc{\hu}{\ensuremath{ H_u}}       
 \newc{\hd}{\ensuremath{ H_d}}

 \newc{\mqthree}{\ensuremath{m_{\widetilde{Q}_3}^2}}
 \newc{\muthree}{\ensuremath{m_{\tilde{u}_3}^2}}
 \newc{\mdthree}{\ensuremath{m_{\tilde{d}_3}^2}}
 \newc{\mlthree}{\ensuremath{m_{\widetilde{L}_3}^2}}
 \newc{\methree}{\ensuremath{m_{\tilde{e}_3}^2}}
 \newc{\mqtwo}{\ensuremath{m_{\widetilde{Q}_2}^2}}
 \newc{\mutwo}{\ensuremath{m_{\tilde{u}_2}^2}}
 \newc{\mdtwo}{\ensuremath{m_{\tilde{d}_2}^2}}
 \newc{\mltwo}{\ensuremath{m_{\widetilde{L}_2}^2}}
 \newc{\metwo}{\ensuremath{m_{\tilde{e}_2}^2}}
 \newc{\mqone}{\ensuremath{m_{\widetilde{Q}_1}^2}}
 \newc{\muone}{\ensuremath{m_{\tilde{u}_1}^2}}
 \newc{\mdone}{\ensuremath{m_{\tilde{d}_1}^2}}
 \newc{\mlone}{\ensuremath{m_{\widetilde{L}_1}^2}}
 \newc{\meone}{\ensuremath{m_{\tilde{e}_1}^2}}
 \newc{\mone}{\ensuremath{M_1}}
 \newc{\monesq}{\ensuremath{M_1^2}}
 \newc{\mtwo}{\ensuremath{M_2}}
 \newc{\mtwosq}{\ensuremath{M_2^2}}
 \newc{\mthree}{\ensuremath{M_3}}
 \newc{\mthreesq}{\ensuremath{M_3^2}}
 \newc{\atau}{\ensuremath{{A_{\tau}}}}
 \newc{\at}{\ensuremath{{A_{t}}}}
 \newc{\ab}{\ensuremath{{A_{b}}}}
 \newc{\atausq}{\ensuremath{{A_{\tau}^2}}}
 \newc{\atsq}{\ensuremath{{A_{t}^2}}}
 \newc{\absq}{\ensuremath{{A_{b}^2}}}
 \newc{\dmzero}{\ensuremath{\Delta{_{m_0}}}}
 \newc{\dmhalf}{\ensuremath{\Delta{_{m_{1/2}}}}}
 \newc{\dmu}{\ensuremath{\Delta{_{\mu}}}}

 \newc{\pten}{\ensuremath{\psi_{10}}}
 \newc{\ffive}{\ensuremath{\phi_{5}}}
 \newc{\hfive}{\ensuremath{h_{5}}}
 \newc{\hbfive}{\ensuremath{h_{\bar{5}}}}
 \newc{\thet}{\ensuremath{\theta_{50}}}
 \newc{\thetb}{\ensuremath{\theta_{\,\overline{50}}}}
 \newc{\ptenhat}{\ensuremath{\hat{\psi}_{10}}}
 \newc{\ffivehat}{\ensuremath{\hat{\phi}_{5}}}
 \newc{\hfivehat}{\ensuremath{\hat{h}_{5}}}
 \newc{\hbfivehat}{\ensuremath{\hat{h}_{\bar{5}}}}
 \newc{\thethat}{\ensuremath{\hat{\theta}_{50}}}
 \newc{\thetbhat}{\ensuremath{\hat{\theta}_{\,\overline{50}}}}
 \newc{\si}{\ensuremath{\Sigma}}
 \newc{\mfive}{\ensuremath{m_5^2}}
 \newc{\mten}{\ensuremath{m_{10}^2}}
 \newc{\dfive}{\ensuremath{\Delta^2_5}}
 \newc{\dbfive}{\ensuremath{\Delta^2_{\bar{5}}}}
 \newc{\dfifty}{\ensuremath{\Delta^2_{50}}}
 \newc{\dfiftyb}{\ensuremath{\Delta^2_{\,\overline{50}}}}
 \newc{\msi}{\ensuremath{m_{\Sigma}^2}}
 \newc{\lamh}{\ensuremath{\lambda_{H}}}
 \newc{\lamhb}{\ensuremath{\lambda_{\bar{H}}}}
 \newc{\ah}{\ensuremath{A_{H}}}
 \newc{\ahb}{\ensuremath{A_{\bar{H}}}}
 \newc{\lams}{\ensuremath{\lambda_{S}}}
 \newc{\as}{\ensuremath{A_{S}}}
 \newc{\lamsig}{\ensuremath{\lambda_{\si}}}
 \newc{\asig}{\ensuremath{A_{\si}}}

 \newc{\msten}{\ensuremath{m_{16}^2}}
 \newc{\mhun}{\ensuremath{m_{126}^2}}
 \newc{\mhunb}{\ensuremath{m_{\bar{126}}^2}}
 \newc{\mthun}{\ensuremath{m_{210}^2}}
 \newc{\ahun}{\ensuremath{A_{\bar{126}}}}
 \newc{\yhun}{\ensuremath{Y_{\bar{126}}}}
 \newc{\aten}{\ensuremath{A_{10}}}
 \newc{\yten}{\ensuremath{Y_{10}}}
 \newc{\alone}{\ensuremath{A_{\lambda_1}}}
 \newc{\altwo}{\ensuremath{A_{\lambda_2}}}
 \newc{\althree}{\ensuremath{A_{\lambda_3}}}
 \newc{\althreeb}{\ensuremath{A_{\bar{\lambda_3}}}}
 \newc{\lone}{\ensuremath{\lambda_1}}
 \newc{\ltwo}{\ensuremath{\lambda_2}}
 \newc{\lthree}{\ensuremath{\lambda_3}}
 \newc{\lthreeb}{\ensuremath{\bar{\lambda_3}}}

\newc{\sigsip}{\ensuremath{\sigma^{\rm SI}_{p}}}	\newc{\sigsin}{\ensuremath{\sigma^{\rm SI}_{n}}}
\newc{\sigsdp}{\ensuremath{\sigma^{\rm SD}_{p}}}	\newc{\sigsdn}{\ensuremath{\sigma^{\rm SD}_{n}}}
\newc{\sigsi}{\ensuremath{\sigma^{\rm SI}}}	\newc{\sigsd}{\ensuremath{\sigma^{\rm SD}}}
\newc{\abund}{\ensuremath{ \Omega h^2}}
\newc{\omegadm}{\ensuremath{ \Omega_{{\rm DM}}}}     \newc{\abunddm}{\ensuremath{ \Omega_{{\rm DM}} h^2}} 
\newc{\omegam}{\ensuremath{ \Omega_{{\rm m}}}}       \newc{\abundm}{\ensuremath{ \Omega_{{\rm m}} h^2}}
\newc{\omegab}{\ensuremath{ \Omega_{{\rm b}}}}	\newc{\abundb}{\ensuremath{ \Omega_{{\rm b}} h^2}}
\newc{\omegatot}{\ensuremath{ \Omega_{{\rm TOT}}}}
\newc{\omegacdm}{\ensuremath{ \Omega_{{\rm CDM}}}}   \newc{\abundcdm}{\ensuremath{ \Omega_{{\rm CDM}} h^2}}
\newc{\omegalambda}{\ensuremath{ \Omega_{\Lambda}}} \newc{\abundlambda}{\ensuremath{ \Omega_{\Lambda} h^2}}
\newc{\omegarad}{\ensuremath{ \Omega_{{\rm rad}}}}  \newc{\abundrad}{\ensuremath{ \Omega_{{\rm rad}} h^2}}
\newc{\rhocrit}{\ensuremath{ \rho_{\rm crit}}}
\newc{\rhochi}{\ensuremath{ \rho_{\chi}}}
\newc{\abunchi}{\ensuremath{\Omega_\chi h^2}}
\newc{\abundlsp}{\ensuremath{\Omega_{\rm LSP}h^2}}
\newc{\amu}{\ensuremath{ a_{\mu}}}        \newc{\amususy}{\ensuremath{ a_{\mu}^{\mathrm{SUSY}}}}
\newc{\amuexpt}{\ensuremath{ a_{\mu}^{\mathrm{expt}}}}        \newc{\amusm}{\ensuremath{ a_{\mu}^{\mathrm{SM}}}}
\newc\deltaamu{\ensuremath{\Delta a_{\mu}}} \newc{\deltaamususy}{\ensuremath{\delta a_{\mu}^{\mathrm{SUSY}}}}
\newc\gmtwo{\ensuremath{ (g-2)_{\mu}}} 
\newc{\deltagmtwomususy}{\ensuremath{\delta\left(g-2\right)_{\mu}^{\mathrm{SUSY}}}}
\newc{\deltagmtwomu}{\ensuremath{\delta\left(g-2\right)_{\mu}}}
\newc\BR{\ensuremath{\rm BR}}
\newc\bsgamma{\ensuremath{ b\rightarrow s \gamma }}
\newc\bxsgamma{\ensuremath{\overline{B}\rightarrow X_{s}\gamma}}
\newc\brbsgamma{\ensuremath{\BR\left(\bsgamma\right)}}
\newc\brbxsgamma{\ensuremath{\BR\left(\bxsgamma\right)}}
\newc\bsmumu{\ensuremath{B_s\to\mu^+\mu^-}}
\newc\brbsmumu{\ensuremath{\BR\left(B_s\to\mu^+\mu^-\right)}}
\newc\bdmmumu{\ensuremath{\overline{B}_d\to\mu^+\mu^-}}
\newc\bbbarmix{\ensuremath{\overline{B}_s\mbox{-}B_s}}      % B_s mixing
\newc\delmbs{\ensuremath{\Delta M_{B_s}}}
\newc{\butaunu}{\ensuremath{B_u \rightarrow \tau \nu}}
\newc{\brbutaunu}{\ensuremath{\BR\left(B_u \rightarrow \tau \nu\right)}}
\let\oldcite\cite
\renewcommand*{\cite}{~\oldcite}

\newcommand*{\hl}{\ensuremath{h}}